\documentclass[aps,pra,twocolumn,showpacs,superscriptaddress,10pt,longbibliography]{revtex4-1}
\usepackage{amssymb}
\usepackage{color}
\usepackage{graphicx}
\usepackage[sort&compress]{natbib}
\usepackage{scalerel}
\usepackage[normalem]{ulem}
\usepackage{graphicx}
\usepackage{xcolor}
\usepackage[colorlinks=true,linkcolor=red,citecolor=blue]{hyperref}
\usepackage{bm}
\usepackage{amsmath}
\usepackage{subfigure}
\usepackage{amsfonts}
\usepackage{mathtools}
\usepackage{braket}
\usepackage{dsfont}
\usepackage{float}
\usepackage{marvosym}
\newcommand{\sect}[1]{\emph{#1}.---}

\begin{document}
\title{Logarithmic expansion of many-body wave packets in random potentials}
 
\author{Arindam Mallick}
\email{marindam@ibs.re.kr}
\affiliation{Center for Theoretical Physics of Complex Systems, Institute for Basic Science(IBS), Daejeon 34126, Korea}
\author{Sergej Flach}
\email{sflach@ibs.re.kr}
\affiliation{Center for Theoretical Physics of Complex Systems, 
Institute for Basic Science(IBS), Daejeon 34126, Korea}

\begin{abstract}
Anderson localization confines the wave function of a quantum particle in a one-dimensional random potential to a volume of the order of the localization length $\xi$. Nonlinear add-ons to the wave dynamics mimic
many-body interactions on a mean field level, and result in escape from the Anderson cage and in unlimited subdiffusion of the interacting cloud.
We address quantum corrections to that subdiffusion by (i) using the ultrafast unitary Floquet dynamics of discrete-time quantum walks, (ii) an interaction strength ramping to 
speed up the subdiffusion, and (iii) an action discretization of the nonlinear terms.
We observe the saturation of the cloud expansion of $N$ particles to a volume $\sim N\xi$. We predict and observe a universal intermediate
logarithmic expansion regime which connects the mean-field diffusion with the final saturation regime and is entirely controlled by 
particle number $N$. 
The temporal window of that regime grows exponentially with the localization length $\xi$. 

\end{abstract}

\maketitle

\sect{Introduction}
Single-particle quantum dynamics in a one-dimensional space with uncorrelated disorder results in Anderson localization (AL), i.e., confinement to 
a finite localization volume of the order of the localization length $\xi$
\cite{Anderson_local_1958_PR}.
The evolution of an initially localized quantum wave packet will consist of an (almost ballistic) expansion up to the volume $ \sim \xi$ \cite{lifshitz1988introduction} with a subsequent halt
and exponential wave-function localization in the tails \cite{Kramer_1993}.
Experimental verifications of AL 
with Bose-Einstein condensates of ultracold atomic gases loaded onto optical potentials were reported harvesting on the halt of the wave-packet expansion
\cite{Billy_2008}.

Many-body interactions alter the picture. Full-scale computations of temporal evolutions are restricted to two or three interacting particles only, with the complexity quickly increasing
due to the Hilbert space dimension proliferation
\cite{shepelyansky1994coherent,imry1995coherent,frahm1995scaling,vonoppen1996interaction,jacquod1997breit,roemer1997no,ponomarev1997coherent,song1997localization,roemer1999two,frahm1999interaction,song1999general,arias1999two,krimer2011two,ivanchenko2014,krimer2015interaction,frahm2016eigenfunction,yusipov2017,Diana_PhysRevB.100.224203,PhysRevB.101.144201_merab}.
Increasing the number of particles
is predicted to result in a slow subdiffusive expansion \cite{Schwiete_PhysRevA.88.053611},
which adds to the computational challenge. The same slow subdiffusion limits experimental studies with condensates due to finite coherence times 
\cite{Lucioni_PhysRevLett.106.230403}.
Treating infinite particle numbers with mean-field approximations results in nonlinear add-ons to the wave dynamics which stem from the two-body interactions. Nonlinear wave-packet expansion
was investigated both analytically and numerically over vast time scales 
\cite{PhysRevLett.70.1787,PhysRevLett.100.094101,flach2009universal,laptyeva2010crossover,PhysRevLett.107.240602,kati2020,ErmannChaos2021}. 
It allows us to obtain the details of the subdiffusion process, with expansion times which are many orders of magnitude larger than the time reached by the experimental implementations of comparable theoretical models \cite{Laptyeva_2014}. Remarkably, the subdiffusive expansion of a nonlinear wave packet appears to show no signatures of halt which was tested using a vast number of different Hamiltonian and discrete-time map evolutions with various types of nonlinear terms \cite{Laptyeva_2014}. 
Contrarily, for quantum clouds with $N$ particles we expect the expansion to stop when the cloud reaches the size of the order of $N\xi$ since each
particle can occupy its own localization volume $\sim\xi$ and is only exponentially weakly interacting with other particles.

Here, we want to explore the long-time wave-packet expansion of an interacting many-body cloud and to establish its slowing down from (sub)diffusion to a complete halt. To achieve that challenging goal, we have to choose proper platforms and approximations. We use a Floquet platform of discrete-time quantum walks which exhibit AL \cite{Ihor_PhysRevB.96.144204_2017}.
Nonlinear add-ons show subdiffusive cloud expansion up to record large evolution times \cite{PhysRevLett.122.040501}.
We further use a time-dependent ramping of the interaction strength which allows us to speed up subdiffusion to normal diffusion \cite{gligoric2013make}.
Finally we quantize the actions in the nonlinear add-ons similar to the Bohr-Sommerfeld quantization approach
 and to previous quantization studies of kicked rotor models \cite{ChirikovGordon1981,BermanPhysicaA1988,BermanChaos1991,GuarneriPRL2014}. As a result, we are able to 
simulate the cloud expansion for tens and hundreds of interacting particles. We carefully choose the localization length $\xi$ and the number of interacting
particles in order to observe the slowing down and halt processes within the time window accessible due to computational restrictions. We succeed in
observing a slowing down of the expansion into a universal intermediate
logarithmic growth regime which connects the (sub)diffusion with the final saturation regime. We derive the analytical details of this
logarithmic regime.

\sect{DTQW} The single-particle linear discrete-time quantum walk (DTQW) is a Floquet evolution of a two-level system $\{\sigma, \bar{\sigma}\}$ on a chain.
The system state at time $t+1$ follows from that at time $t$ by the following unitary map:
\begin{align}
 \psi_{n,\sigma}(t+1) &= \cos \theta \psi_{n-1,\sigma}(t) +  e^{i\phi_{n-1}(t)}\sin \theta \psi_{n-1,\bar{\sigma}}(t),\notag\\
 \psi_{n,\bar{\sigma}}(t+1) &= -e^{-i\phi_{n+1}(t)}\sin \theta \psi_{n+1,\sigma}(t) +  \cos \theta \psi_{n+1,\bar{\sigma}}(t),
\end{align}
where $n$ counts the lattice sites, and $\theta$ is the mixing angle in the $\{\sigma, \bar{\sigma}\}$ space which controls the kinetic energy of an excitation \cite{Ihor_PhysRevB.96.144204_2017}.
The uncorrelated random on-site disorder in the phase $\phi_n\equiv \zeta_n  \in [-\pi, \pi]$ results in Anderson localization with the localization length $\xi = -[\ln(|\cos \theta|)]^{-1}$ \cite{Ihor_PhysRevB.96.144204_2017}.  Note that $\xi(\theta \rightarrow 0) \rightarrow \infty$ and $\xi(\theta \rightarrow \pi/2) \rightarrow 0$.
DTQWs were introduced as a quantum version of
classical random walks \cite{Aharonov_PhysRevA.48.1687}. They serve as a single particle version of a quantum cellular automaton \cite{Meyer_1996, Mallick_2016}. DTQWs became a useful tool to study various systems and phenomena such as relativistic particles, artificial gauge fields \cite{mallick2019quantum, Arnault_2016}, various topological phases \cite{Kitagawa_2010, Asboth_PhysRevB.86.195414, Asboth_PhysRevB.102.224202}, percolation problems \cite{chandrashekar2014quantum, Koll_r_2014}, localization phenomena \cite{Koll_r_2020}, and implementation of quantum information tasks \cite{chawla2020discrete, Srikara_2020, Vlachou_2018}, among others. DTQW were implemented experimentally using NMR devices \cite{Ryan_2005}, optical devices \cite{Crespi_2013, Su_2019}, in the IBM quantum computer \cite{Acasiete_2020}, and in a trapped ion quantum computer \cite{Huerta_Alderete_2020}.  

Nonlinearity was introduced in DTQWs in a number of publications \cite{Navarrete_PhysRevA.75.062333,PhysRevLett.122.040501,Maeda_2019, maeda2020continuous, mochizuki2020stability, adami2019nonlinear}. We follow 
Ref.~\cite{PhysRevLett.122.040501} by making the phase $\phi_n(t)$ a continuous function of the local norm $\rho_n(t)= |\psi_{n, \sigma}(t)|^2 + |\psi_{n, \bar{\sigma}}(t)|^2$:
\begin{equation}
\phi_n = \gamma \rho_n +\zeta_n \;.
\label{nlphase}
\end{equation}
Both linear and nonlinear DTQWs preserve the total norm $\mathcal{A}=\sum_n \rho_n$.
We will evolve the DTQWs starting with one and the same localized initial condition $[\psi_{n, \sigma}(0), \psi_{n, \bar{\sigma}}(0)]$ =  $\delta_{n,0}[1, i]/\sqrt{2}$ with $\mathcal{A}=1$. 
For $\gamma=0$ the corresponding linear DTQW results in a short expansion and final halt of the wave packet spreading due to Anderson localization.
Instead, for $\gamma \neq 0$ the wave packet continues its expansion beyond the limits set by Anderson localization \cite{PhysRevLett.122.040501}. Its root mean square (rms) 
\begin{align}
 r = \sqrt{\braket{n^2} - \braket{n}^2}\; , \; \braket{n^x} = \sum_n n^x \rho_n,
\end{align}
grows indefinitely in a subdiffusive manner $r(t)\sim t^{1/6}$  \cite{PhysRevLett.122.040501}.

\begin{figure}
 \includegraphics[width = 8.5cm]{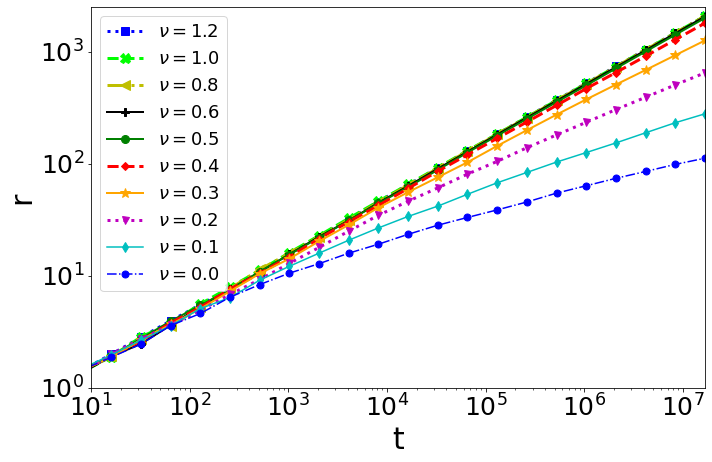}
 \caption{Average of $\log r$ versus $\log t$ for expanding wave packets with ramping nonlinearity.
The ramping exponent $\nu=0, 0.1, 0.2, 0.3, 0.4, 0.5, 0.6, 0.8, 1.0, 1.2$ increases from bottom to top.
 Here, $\theta = 0.35\pi$ and $\gamma=30$. Averaging is performed over 24 random disorder realizations.}
\label{fig:nonl_qw}
\end{figure}

\sect{Ramping}
The subdiffusive expansion is a rather slow process since it is characterized by a time-dependent diffusion constant which is 
a function of the wave packet
 norm density $\rho \sim 1/r$: $D\equiv D(\gamma \rho)$ \cite{FLACH2010548}. The density $\rho$ is
decaying in time while the wave packet expands, thus effectively slowing down the diffusion. Since we intend to simulate the impact of a yet to be introduced quantum slowing down correction, we
are facing a challenging computational task. Remarkably, there is a reported way to speed up the subdiffusive process by choosing a proper ramping of the strength of nonlinear interaction $\gamma(t)$.
Such a ramping intends to compensate for the decrease of the density $\rho$ through a proper increase of the interaction strength $\gamma$. That ramping can be in principle
realized in experiments with ultracold atoms through a time-dependent magnetic field which controls the two-body scattering length in a vicinity to Feshbach resonances 
\cite{inouye1998observation, khaykovich2002formation, weber2003bose, Roati_PhysRevLett.99.010403_2007}.
 The ramping speedup scheme was successfully tested with a discrete nonlinear Schr\"odinger lattice Hamiltonian and a nonlinear quantum kicked rotor map \cite{gligoric2013make}.
We follow the ramping protocol from Ref.~\cite{gligoric2013make} and choose $\gamma(t) = \gamma t^\nu$:
\begin{equation}
\phi_n(t) = \gamma t^\nu  \rho_n(t)  + \zeta_n \label{eq:nl} \; .
\end{equation}
In a one dimensional diffusive process for an initially localized wave packet, its variance grows linearly in time: $r^2 = D t$.
 For the ramping case normal diffusion is reached when $D$ is constant (stationary), implying that
the product $(\gamma t^\nu \rho)$ inside the wave packet stays approximately constant. 
Since $\rho\sim 1/r$ and $r\sim \sqrt{t}$ for normal diffusion, we conclude that the ramping exponent $\nu =1/2$ ensures normal diffusion
of the wave packet (see also Ref.~\cite{gligoric2013make}). A further increase of the ramping exponent beyond $1/2$ does not modify the obtained normal diffusion \cite{FLACH2010548}.
Smaller ramping exponents gradually slow down the spreading into subdiffusion.
Our numerical results in Fig.~\ref{fig:nonl_qw} confirm the above considerations. For what follows we will use $\nu=1$ and $\gamma=30$.

\sect{Mimicking quantization through discretization}
We arrived at the final and central part of our complex evolution design which intends to mimic a finite number of interacting
quantum particles. 
The quantum analog of the total norm $\mathcal{A}$ is the number of particles $N$ similar to the relation between the total norm of a Gross-Pitaevskii equation and the number of particles in a corresponding Bose-Hubbard Hamiltonian \cite{Dalfovo_RevModPhys_1999, Smerzi_PRA_2003}. The quantum analog of the norm density $\rho_n$ is the number of particles on that site. In analogy to the particle-number-dependent interaction energy of a many body quantum lattice model
we discretize the density $\rho_n$ inside the nonlinear term [\eqref{nlphase} and \eqref{eq:nl}] using a step function to arrive at
 \begin{align}
\phi_n(t) = \frac{\gamma t^\nu}{N} \left\lfloor N \rho_n(t)\right\rfloor  + \zeta_n, \label{q_nl}
\end{align}
 where $\left\lfloor N \rho_n(t)\right\rfloor$ is the largest possible integer less than or equal to $N \rho_n(t)$. The parameter $N \geq 1$ 
serves as the analog of the number of particles in a quantum many-body system. 
Note that the total number of particles $\sum_n N \rho_n(t)$ = $N$ is conserved.
For $N \to \infty$ we recover the continuous density dependence of the phase (\ref{eq:nl}).
\begin{figure}
\includegraphics[width = 9.3cm]{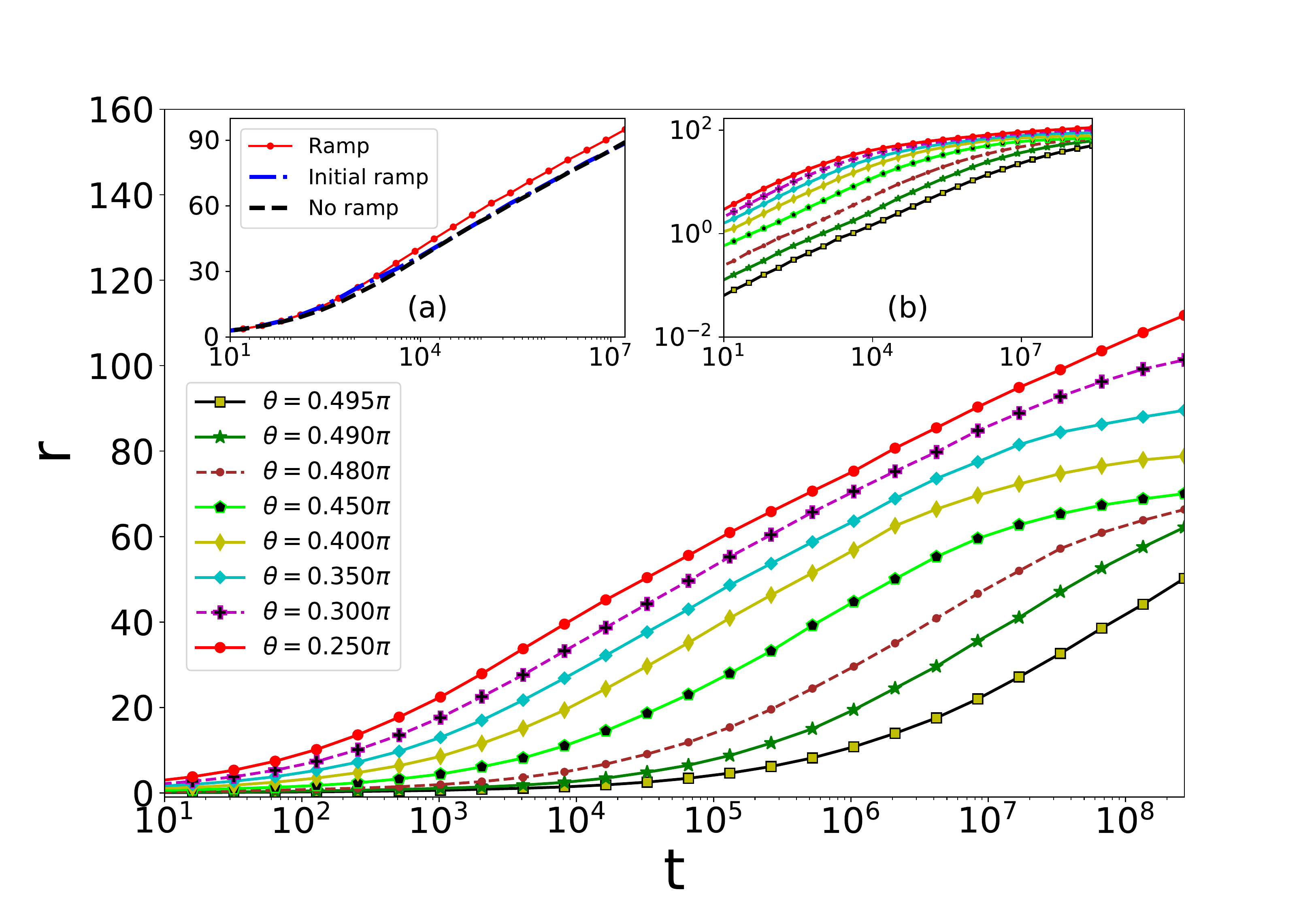}
 \caption{Average of $r$ vs $\log t$ for various values of $\theta$ (numbers in legend, increasing from top to bottom)  and
$N = 24$. The average is taken over 177 random realizations. 
Inset (a) left: Average of $r$ vs $\log t$ for $\theta = 0.25 \pi$ and with ramping, only initial ramping, and no ramping (see text for details).
Inset (b) right: the same main plot but in log-log scale.
Averages are now taken for $\log r$.}
 \label{fig:m2_N_24}
\end{figure}

 Figure \ref{fig:m2_N_24}{\color{red}(b)} shows the computed dependence $\log r$ vs $\log t$ for $N=24$, $\gamma=30$, and $\nu=1$ and
a variety of different angles $\theta$ which control the single-particle localization length $\xi = -[\ln(|\cos \theta|)]^{-1}$. 
The wave packet initially expands diffusively and shows clear signatures of saturation and halt at larger evolution times. The rms value $r$
at saturation can be expected to be of the order of the number of particles $N$ times the volume of a one-particle Anderson localized 
wave packet $v_l$. That single-particle volume will depend on the localization length $\xi$ and we will assess these details further below.
Interestingly, Fig.~\ref{fig:m2_N_24}{\color{red}(b)} appears to predict that the saturation and halt will happen at earlier times the smaller
the angle $\theta$ and therefore the larger the localization length $\xi$. However, a replot of the same data with $r$ on a linear scale
in the main part of Fig.~\ref{fig:m2_N_24} shows that what appeared to be a saturation and halt on logarithmic rms scales, turns
into a {\sl logarithmic expansion regime} of the quantized wave packet,
\begin{align}
 r(t) = r_0 + D_{\text{qn}} \log_{10} (t), \label{eq:log_gr}
\end{align}
where $r_0$ is a fitting parameter which will be quantified below. It can easily take negative values.
The smaller $\theta$ and the larger $\xi$, the earlier the
logarithmic expansion set in, and the further it extends in time.
 In Fig.~\ref{fig:m2_N_24}{\color{red}(a)} we replot the data for $\theta = 0.25 \pi$ and compare
with a run where ramping is switched off when the rms reaches the value $r=25$ where the logarithmic spreading
regime appears to start. We also plot data from a run where no ramping is applied altogether. We observe
very good agreement between all curves, which clearly shows that the ramping protocol is not affecting the essential details of the logarithmic spreading. The ramping though is crucial for larger $\theta$ values in order to faster reach
the onset of logarithmic spreading.

\begin{figure}
 \includegraphics[width = 8.5cm]{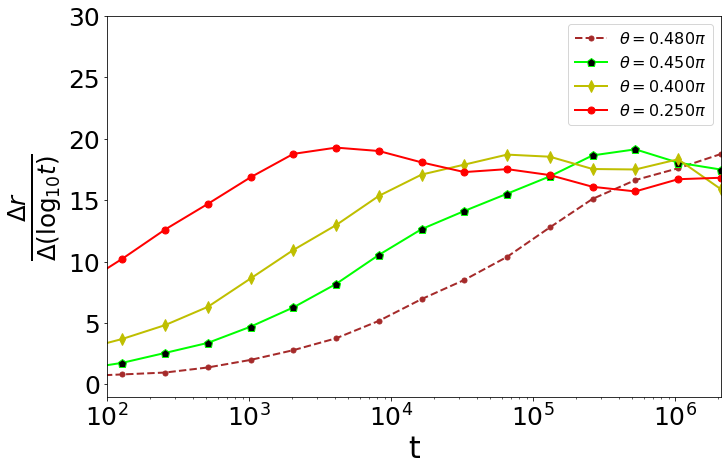}
 \caption{Derivative of selected curves in Fig.~\ref{fig:m2_N_24} vs $\log t$. Only four $\theta$ values are shown for the sake of clarity. Curves for other $\theta$ values show similar behavior.
}
 \label{fig:slope_rms_log}
\end{figure}

\begin{figure}
\includegraphics[width = 8.5cm]{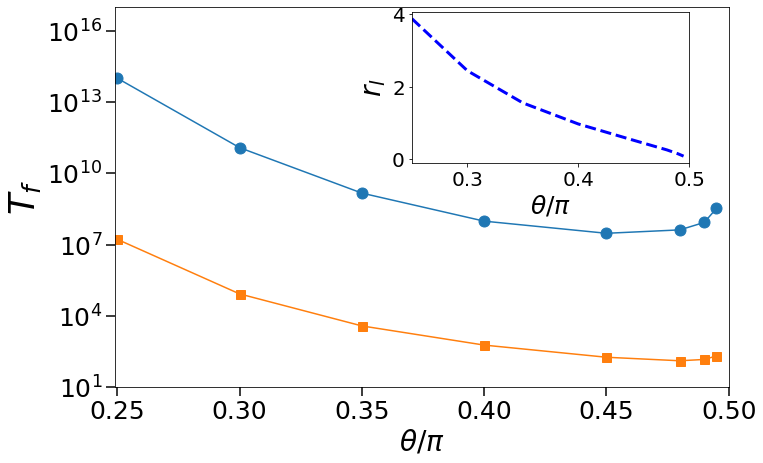}
\caption{
$T_f$ versus $\theta$. Blue top curve data are extracted from the data in Fig.~\ref{fig:m2_N_24} following Eq.~(\ref{eq:loc_vol}). For the fit details, see the text.
Orange bottom curve---result of theoretical analysis.
Inset: The saturated rms $r_l$ for a single particle as a function of $\theta$.}
\label{fig:sat_time}
\end{figure}

To further substantiate this finding, we compute the local derivatives of the curves from the main panel of Fig.~\ref{fig:m2_N_24} and plot them
in Fig.~\ref{fig:slope_rms_log}. We find that the derivatives show a plateau-like structure in the regime of logarithmic expansion, with
 a slope value $D_\text{qn} \approx 20$, almost independently of $\theta$ and $\xi$.
Logarithmic numerical derivatives are notorious for their fluctuations due to finite numbers of disorder realizations and smoothening operations resulting in slow fluctuations, therefore we will not analyze possible fine structures.
Let us measure the true saturation time $T_f$. For that we 
find the largest slope position and value in Fig.~\ref{fig:slope_rms_log}. We then use the linear fit (\ref{eq:log_gr})
and extend it to larger times until the rms reaches the assumed final localization volume
$V_\text{loc}$. The read-off time is identified as $T_f$:
\begin{align} 
 T_f = 10^{(V_\text{loc} - r_{_0})/D_{\text{qn}}}\;. \label{eq:loc_vol}
\end{align}
To estimate $V_\text{loc}$ we assume that it is proportional to the number of particles $N$
and the single-particle localization volume $v_l\approx r_l+1$ where $r_l\sim \xi$ is the saturated rms of a single particle: 
\begin{align}
 V_\text{loc} = N f (r_l + 1)\;.
\label{Vloc}
\end{align}
Note that the correction by one integer in (\ref{Vloc}) accounts for the case of small localization length $\xi \ll 1$ when $r_l \ll 1$ but the volume
is approximately one lattice site.
We plot the value of $r_l$ as a function of $\theta$ in the inset of Fig.~\ref{fig:sat_time}.
The proportionality factor $f$ can be assumed to be of order one. To get a number, we use the data for $\theta=0.45 \pi$ in Fig.~\ref{fig:m2_N_24}
to read off $V_\text{loc} \approx 70$ and arrive at the value $f\approx 2$ which we will use for all other curve analysis as well. The outcome is
shown by the blue (top) curve in Fig.~\ref{fig:sat_time}. We find that $T_f(\theta)$ is expected to have a minimum at around $\theta=0.45 \pi$,
while it is growing substantially when deviating to larger and smaller values of $\theta$.

\sect{Derivation}
In order to explain the observed logarithmic expansion and the subsequent halt, we assume that the localization length $\xi \gg 1$ and thus $r_l \gg 1$. 
In order to enter the logarithmic expansion regime with $r(t) > r_l$, the wave-packet density must be small such that at any given time for most of the wave-packet sites
the quantization condition (\ref{q_nl}) yields one particle, i.e., $\left\lfloor N \rho \right\rfloor = 0$.
For a large time the dynamics in a localization volume $v_l$ will be following the linear DTQW dynamics. We assume (and tested numerically) that
the local norm on any of the two levels
in a two-level system will fluctuate
following an exponential (Gibbs) distribution  $\sim  e^{-\mu \rho}$ with chemical potential $\mu=2r$ (inverse of average local norm). The fluctuations are restricted to norm redistributions within 
the finite volume $v_l\approx r_l$ and have an upper limit $\rho_\text{max} = c r_l / r(t) > 1/N$ with the constant $c$ being of order one and to be fixed below. 
Then the norm $\rho_n$ on the entire two-level system will fluctuate according to the distribution of the sum of two uncorrelated non-negative random numbers
\begin{equation}
\mathcal{P}(\rho_n)=\mu^2 \rho_n {\rm e}^{-\mu \rho_n}\;,\; 0 \leq \rho_n \leq \rho_\text{max}\;. \label{gibbs}
\end{equation}
There is a small but finite probability $\wp$ that $1/N < \rho_n < \rho_\text{max}$:
\begin{equation}
\wp (r) = \int_{1/N}^{\rho_{max}} \mathcal{P} d\rho_n \approx e^{-2r/N} - e^{- 2cr_l } \;.
\label{probability}
\end{equation}
The average time for that to happen is $T \approx 1/\wp$. Once the rare event takes place, the quantized nonlinearity (\ref{q_nl}) effectively changes
the disorder potential within the considered localization volume. At the edge of the wave packet, that leads to an expansion into a newly accessible
localization volume: 
\begin{equation}
\frac{dr}{dt} = r_l \wp(r)\;,\; r(t) = N c r_l + \frac{N}{2} \ln\left(1 - e^{- \frac{2 r_l t}{N e^{2 c r_l}}}\right) \;.
\label{ode}
\end{equation}
It follows that $r(t \rightarrow \infty) \rightarrow N c r_l $ and with (\ref{Vloc}) we conclude $ c \approx f$. We are now in position to obtain the logarithmic
expansion law, which is derived from (\ref{ode}):
\begin{equation}
r(t) = \frac{N}{2} \ln\left(\frac{2 r_l}{N}\right) +  \frac{N}{2} \ln\left(t\right) \;,\; t \ll \frac{N e^{2f r_l}}{2 r_l} \;.
\label{lnlaw}
\end{equation}
The fitting parameter $r_0$ from Eq.~(\ref{eq:log_gr}) corresponds to the first term on the right-hand side (rhs) of (\ref{lnlaw}).
The logarithmic expansion is universal in the sense that the slope $N/2$ depends only on the particle number, but not on the particularities
of the disorder, nonlinearity, and ramping. In our numerical computations in Fig.~\ref{fig:slope_rms_log} we used $N=24$. The observed slope
$\approx 20$ was measured on logarithmic timescales in base 10 and matches reasonably well with the theoretical prediction $\frac{N}{2} \ln 10 = 27.6$.
Finally we can attempt to estimate the saturation time $T_f$ from (\ref{ode}). For that we compute the time at which the exponent in the rhs
in (\ref{ode}) is of order one. The result reads
\begin{equation}
T_f = \frac{N e^{2f r_l}}{2 r_l} \;.
\label{tf}
\end{equation}
It follows that the saturation time diverges exponentially with large $r_l$ in the limit of infinite localization length $\theta \rightarrow 0$. At the same time the saturation 
time diverges as $1/r_l$ in the limit of small localization length $\theta \rightarrow \pi/2$. Therefore there must be a minimum saturation time
$T_f$ at some value of $\theta$. The full dependence of $T_f(\theta)$ is plotted in Fig.~\ref{fig:sat_time}. We find that the minimum saturation time
is obtained for $\theta \approx 0.48 \pi$ which is reasonably close to the numerically observed value $\approx 0.45 \pi$.
Note that the absolute values of the theoretical estimate of $T_f$ are much lower than the computational results. The reason is that our theoretical approach
is qualitative when it comes to fitting,  and operates on logarithmic scales. Changing $V_\text{loc}$ from 70 to around 80 changes $f$ from 2 to 2.3. Additional replacement of $r_l$ by
$r_l+1$ results in a magnitude shift of $T_f$ upwards by three orders of magnitude. Despite the qualitative character of the theory, it is 
capable of reproducing the universal log law, and the minimum in $T_f$.

\sect{Discussion}\label{disc_sec} The wave packet starts to spread in a ballistic-like regime up to the size of the
single particle localization volume $v_l$. It then enters a subdiffusive regime when not ramped, or diffusive regime when ramped which extends until the start of the logarithmic regime due to density quantization which happens for a wave-packet size of the order of $N$ and at a time $t \sim N^2/D$. The spreading finally halts for a packet size of the order of $N v_l$. Therefore both the logarithmic part and the halt are scaled to infinite times in the limit $N\rightarrow \infty$, which recovers familiar
(sub)diffusion of the wave packet on all accessible time scales known for the nonquantized theory \cite{PhysRevLett.122.040501}. If instead the single-particle localization volume $v_l$ is tuned to larger and larger values, the logarithmic spreading part is extending over more and more time decades. 
The crossover between the (sub)diffusion and the logarithmic regimes is scaled to shorter times. This follows
from the dependence of the diffusion constant $D$ on the localization length \cite{Laptyeva_2014}.
The (sub)diffusive regime window is closing completely for $v_l > N$ so that the ballistic regime is immediately followed by the logarithmic one.
All these results can be read off the data in Fig.~\ref{fig:m2_N_24}. 

Experimental platforms which use ultracold atomic gases \cite{Lucioni_PhysRevLett.106.230403} can keep cloud coherence up to times which are
comparable to $10^4$ of our dimensionless time units  \cite{Laptyeva_2014}. Interaction strength ramping
using Feshbach resonances is feasible. Therefore we conclude that such experimental platforms can observe 
the onset of logarithmic spreading (see the curve for $\theta=0.25 \pi$ in Fig.~\ref{fig:slope_rms_log}).

 \sect{Conclusion}\label{concl_sec}
Nonlinear wave packets spread subdiffusively in a disordered environment. Despite many efforts to observe a slowing down of the subdiffusion, all
computational evidence points to unlimited subdiffusion. Quantum systems with many particles and conserved particle number were expected
to be sufficient for a slowing down from subdiffusion and final halt of spreading for finite particle numbers. In this Letter we simulate this effect
using a rough action quantization of a nonlinear wave propagation. To actually reach the desired timescales, we chose highly efficient unitary maps
(discrete-time quantum walks) and nonlinear interaction strength ramping for computational speed up. We succeeded with observing the halt of expansion.
In addition, we discovered an intermediate logarithmic expansion regime whose time window grows with increasing localization length.
In that regime the speed of the wave-packet size growth on logarithmic timescales depends only on the total particle number in the packet.

\sect{Acknowledgements}
We thank M. Malishava and I. Vakulchyk for helpful discussions. This  work  was  supported  by  the  Institute  for  Basic Science in Korea (IBS-R024-D1).

 \bibliography{Qn_nl_qw,mbl}

\providecommand{\noopsort}[1]{}\providecommand{\singleletter}[1]{#1}%
\begin{thebibliography}{71}%
\makeatletter
\providecommand \@ifxundefined [1]{%
 \@ifx{#1\undefined}
}%
\providecommand \@ifnum [1]{%
 \ifnum #1\expandafter \@firstoftwo
 \else \expandafter \@secondoftwo
 \fi
}%
\providecommand \@ifx [1]{%
 \ifx #1\expandafter \@firstoftwo
 \else \expandafter \@secondoftwo
 \fi
}%
\providecommand \natexlab [1]{#1}%
\providecommand \enquote  [1]{``#1''}%
\providecommand \bibnamefont  [1]{#1}%
\providecommand \bibfnamefont [1]{#1}%
\providecommand \citenamefont [1]{#1}%
\providecommand \href@noop [0]{\@secondoftwo}%
\providecommand \href [0]{\begingroup \@sanitize@url \@href}%
\providecommand \@href[1]{\@@startlink{#1}\@@href}%
\providecommand \@@href[1]{\endgroup#1\@@endlink}%
\providecommand \@sanitize@url [0]{\catcode `\\12\catcode `\$12\catcode
  `\&12\catcode `\#12\catcode `\^12\catcode `\_12\catcode `\%12\relax}%
\providecommand \@@startlink[1]{}%
\providecommand \@@endlink[0]{}%
\providecommand \url  [0]{\begingroup\@sanitize@url \@url }%
\providecommand \@url [1]{\endgroup\@href {#1}{\urlprefix }}%
\providecommand \urlprefix  [0]{URL }%
\providecommand \Eprint [0]{\href }%
\providecommand \doibase [0]{http://dx.doi.org/}%
\providecommand \selectlanguage [0]{\@gobble}%
\providecommand \bibinfo  [0]{\@secondoftwo}%
\providecommand \bibfield  [0]{\@secondoftwo}%
\providecommand \translation [1]{[#1]}%
\providecommand \BibitemOpen [0]{}%
\providecommand \bibitemStop [0]{}%
\providecommand \bibitemNoStop [0]{.\EOS\space}%
\providecommand \EOS [0]{\spacefactor3000\relax}%
\providecommand \BibitemShut  [1]{\csname bibitem#1\endcsname}%
\let\auto@bib@innerbib\@empty
\bibitem [{\citenamefont {Anderson}(1958)}]{Anderson_local_1958_PR}%
  \BibitemOpen
  \bibfield  {author} {\bibinfo {author} {\bibfnamefont {P.~W.}\ \bibnamefont
  {Anderson}},\ }\bibfield  {title} {\enquote {\bibinfo {title} {Absence of
  diffusion in certain random lattices},}\ }\href {\doibase
  10.1103/PhysRev.109.1492} {\bibfield  {journal} {\bibinfo  {journal} {Phys.
  Rev.}\ }\textbf {\bibinfo {volume} {109}},\ \bibinfo {pages} {1492--1505}
  (\bibinfo {year} {1958})}\BibitemShut {NoStop}%
\bibitem [{\citenamefont {Lifshitz}\ \emph {et~al.}()\citenamefont {Lifshitz},
  \citenamefont {Gredeskul},\ and\ \citenamefont
  {Pastur}}]{lifshitz1988introduction}%
  \BibitemOpen
  \bibfield  {author} {\bibinfo {author} {\bibfnamefont {I.~M.}\ \bibnamefont
  {Lifshitz}}, \bibinfo {author} {\bibfnamefont {S.~A.}\ \bibnamefont
  {Gredeskul}}, \ and\ \bibinfo {author} {\bibfnamefont {L.~A.}\ \bibnamefont
  {Pastur}},\ }\bibfield  {title} {\enquote {\bibinfo {title} {Introduction to
  the theory of disordered systems},}\ }\href@noop {} {\bibinfo  {journal}
  {(Wiley, New York 1988)}\ }\BibitemShut {NoStop}%
\bibitem [{\citenamefont {Kramer}\ and\ \citenamefont
  {MacKinnon}(1993)}]{Kramer_1993}%
  \BibitemOpen
\bibfield  {journal} {  }\bibfield  {author} {\bibinfo {author} {\bibfnamefont
  {B}~\bibnamefont {Kramer}}\ and\ \bibinfo {author} {\bibfnamefont
  {A}~\bibnamefont {MacKinnon}},\ }\bibfield  {title} {\enquote {\bibinfo
  {title} {Localization: theory and experiment},}\ }\href {\doibase
  10.1088/0034-4885/56/12/001} {\bibfield  {journal} {\bibinfo  {journal}
  {Reports on Progress in Physics}\ }\textbf {\bibinfo {volume} {56}},\
  \bibinfo {pages} {1469--1564} (\bibinfo {year} {1993})}\BibitemShut {NoStop}%
\bibitem [{\citenamefont {Billy}\ \emph {et~al.}(2008)\citenamefont {Billy},
  \citenamefont {Josse}, \citenamefont {Zuo}, \citenamefont {Bernard},
  \citenamefont {Hambrecht}, \citenamefont {Lugan}, \citenamefont
  {Cl{\'e}ment}, \citenamefont {Sanchez-Palencia}, \citenamefont {Bouyer},\
  and\ \citenamefont {Aspect}}]{Billy_2008}%
  \BibitemOpen
  \bibfield  {author} {\bibinfo {author} {\bibfnamefont {Juliette}\
  \bibnamefont {Billy}}, \bibinfo {author} {\bibfnamefont {Vincent}\
  \bibnamefont {Josse}}, \bibinfo {author} {\bibfnamefont {Zhanchun}\
  \bibnamefont {Zuo}}, \bibinfo {author} {\bibfnamefont {Alain}\ \bibnamefont
  {Bernard}}, \bibinfo {author} {\bibfnamefont {Ben}\ \bibnamefont
  {Hambrecht}}, \bibinfo {author} {\bibfnamefont {Pierre}\ \bibnamefont
  {Lugan}}, \bibinfo {author} {\bibfnamefont {David}\ \bibnamefont
  {Cl{\'e}ment}}, \bibinfo {author} {\bibfnamefont {Laurent}\ \bibnamefont
  {Sanchez-Palencia}}, \bibinfo {author} {\bibfnamefont {Philippe}\
  \bibnamefont {Bouyer}}, \ and\ \bibinfo {author} {\bibfnamefont {Alain}\
  \bibnamefont {Aspect}},\ }\bibfield  {title} {\enquote {\bibinfo {title}
  {Direct observation of anderson localization of matter waves in a controlled
  disorder},}\ }\href {\doibase 10.1038/nature07000} {\bibfield  {journal}
  {\bibinfo  {journal} {Nature}\ }\textbf {\bibinfo {volume} {453}},\ \bibinfo
  {pages} {891--894} (\bibinfo {year} {2008})}\BibitemShut {NoStop}%
\bibitem [{\citenamefont {Shepelyansky}(1994)}]{shepelyansky1994coherent}%
  \BibitemOpen
  \bibfield  {author} {\bibinfo {author} {\bibfnamefont {D.~L.}\ \bibnamefont
  {Shepelyansky}},\ }\bibfield  {title} {\enquote {\bibinfo {title} {Coherent
  propagation of two interacting particles in a random potential},}\ }\href
  {\doibase 10.1103/PhysRevLett.73.2607} {\bibfield  {journal} {\bibinfo
  {journal} {Phys. Rev. Lett.}\ }\textbf {\bibinfo {volume} {73}},\ \bibinfo
  {pages} {2607--2610} (\bibinfo {year} {1994})}\BibitemShut {NoStop}%
\bibitem [{\citenamefont {Imry}(1995)}]{imry1995coherent}%
  \BibitemOpen
  \bibfield  {author} {\bibinfo {author} {\bibfnamefont {Y}~\bibnamefont
  {Imry}},\ }\bibfield  {title} {\enquote {\bibinfo {title} {Coherent
  propagation of two interacting particles in a random potential},}\ }\href
  {\doibase 10.1209/0295-5075/30/7/005} {\bibfield  {journal} {\bibinfo
  {journal} {Europhys. Lett. ({EPL})}\ }\textbf {\bibinfo {volume} {30}},\
  \bibinfo {pages} {405--408} (\bibinfo {year} {1995})}\BibitemShut {NoStop}%
\bibitem [{\citenamefont {Frahm}\ \emph {et~al.}(1995)\citenamefont {Frahm},
  \citenamefont {Müller-Groeling}, \citenamefont {Pichard},\ and\
  \citenamefont {Weinmann}}]{frahm1995scaling}%
  \BibitemOpen
  \bibfield  {author} {\bibinfo {author} {\bibfnamefont {K}~\bibnamefont
  {Frahm}}, \bibinfo {author} {\bibfnamefont {A}~\bibnamefont
  {Müller-Groeling}}, \bibinfo {author} {\bibfnamefont {J.-L}\ \bibnamefont
  {Pichard}}, \ and\ \bibinfo {author} {\bibfnamefont {D}~\bibnamefont
  {Weinmann}},\ }\bibfield  {title} {\enquote {\bibinfo {title} {Scaling in
  interaction-assisted coherent transport},}\ }\href {\doibase
  10.1209/0295-5075/31/3/008} {\bibfield  {journal} {\bibinfo  {journal}
  {Europhys. Lett. ({EPL})}\ }\textbf {\bibinfo {volume} {31}},\ \bibinfo
  {pages} {169--174} (\bibinfo {year} {1995})}\BibitemShut {NoStop}%
\bibitem [{\citenamefont {von Oppen}\ \emph {et~al.}(1996)\citenamefont {von
  Oppen}, \citenamefont {Wettig},\ and\ \citenamefont
  {M\"uller}}]{vonoppen1996interaction}%
  \BibitemOpen
  \bibfield  {author} {\bibinfo {author} {\bibfnamefont {Felix}\ \bibnamefont
  {von Oppen}}, \bibinfo {author} {\bibfnamefont {Tilo}\ \bibnamefont
  {Wettig}}, \ and\ \bibinfo {author} {\bibfnamefont {Jochen}\ \bibnamefont
  {M\"uller}},\ }\bibfield  {title} {\enquote {\bibinfo {title}
  {Interaction-induced delocalization of two particles in a random potential:
  Scaling properties},}\ }\href {\doibase 10.1103/PhysRevLett.76.491}
  {\bibfield  {journal} {\bibinfo  {journal} {Phys. Rev. Lett.}\ }\textbf
  {\bibinfo {volume} {76}},\ \bibinfo {pages} {491--494} (\bibinfo {year}
  {1996})}\BibitemShut {NoStop}%
\bibitem [{\citenamefont {Jacquod}\ \emph {et~al.}(1997)\citenamefont
  {Jacquod}, \citenamefont {Shepelyansky},\ and\ \citenamefont
  {Sushkov}}]{jacquod1997breit}%
  \BibitemOpen
  \bibfield  {author} {\bibinfo {author} {\bibfnamefont {Ph.}\ \bibnamefont
  {Jacquod}}, \bibinfo {author} {\bibfnamefont {D.~L.}\ \bibnamefont
  {Shepelyansky}}, \ and\ \bibinfo {author} {\bibfnamefont {O.~P.}\
  \bibnamefont {Sushkov}},\ }\bibfield  {title} {\enquote {\bibinfo {title}
  {Breit-wigner width for two interacting particles in a one-dimensional random
  potential},}\ }\href {\doibase 10.1103/PhysRevLett.78.923} {\bibfield
  {journal} {\bibinfo  {journal} {Phys. Rev. Lett.}\ }\textbf {\bibinfo
  {volume} {78}},\ \bibinfo {pages} {923--926} (\bibinfo {year}
  {1997})}\BibitemShut {NoStop}%
\bibitem [{\citenamefont {R\"omer}\ and\ \citenamefont
  {Schreiber}(1997)}]{roemer1997no}%
  \BibitemOpen
  \bibfield  {author} {\bibinfo {author} {\bibfnamefont {Rudolf~A.}\
  \bibnamefont {R\"omer}}\ and\ \bibinfo {author} {\bibfnamefont {Michael}\
  \bibnamefont {Schreiber}},\ }\bibfield  {title} {\enquote {\bibinfo {title}
  {No enhancement of the localization length for two interacting particles in a
  random potential},}\ }\href {\doibase 10.1103/PhysRevLett.78.515} {\bibfield
  {journal} {\bibinfo  {journal} {Phys. Rev. Lett.}\ }\textbf {\bibinfo
  {volume} {78}},\ \bibinfo {pages} {515--518} (\bibinfo {year}
  {1997})}\BibitemShut {NoStop}%
\bibitem [{\citenamefont {Ponomarev}\ and\ \citenamefont
  {Silvestrov}(1997)}]{ponomarev1997coherent}%
  \BibitemOpen
  \bibfield  {author} {\bibinfo {author} {\bibfnamefont {I.~V.}\ \bibnamefont
  {Ponomarev}}\ and\ \bibinfo {author} {\bibfnamefont {P.~G.}\ \bibnamefont
  {Silvestrov}},\ }\bibfield  {title} {\enquote {\bibinfo {title} {Coherent
  propagation of interacting particles in a random potential: The mechanism of
  enhancement},}\ }\href {\doibase 10.1103/PhysRevB.56.3742} {\bibfield
  {journal} {\bibinfo  {journal} {Phys. Rev. B}\ }\textbf {\bibinfo {volume}
  {56}},\ \bibinfo {pages} {3742--3759} (\bibinfo {year} {1997})}\BibitemShut
  {NoStop}%
\bibitem [{\citenamefont {Song}\ and\ \citenamefont
  {Kim}(1997)}]{song1997localization}%
  \BibitemOpen
  \bibfield  {author} {\bibinfo {author} {\bibfnamefont {P.~H.}\ \bibnamefont
  {Song}}\ and\ \bibinfo {author} {\bibfnamefont {Doochul}\ \bibnamefont
  {Kim}},\ }\bibfield  {title} {\enquote {\bibinfo {title} {Localization of two
  interacting particles in a one-dimensional random potential},}\ }\href
  {\doibase 10.1103/PhysRevB.56.12217} {\bibfield  {journal} {\bibinfo
  {journal} {Phys. Rev. B}\ }\textbf {\bibinfo {volume} {56}},\ \bibinfo
  {pages} {12217--12220} (\bibinfo {year} {1997})}\BibitemShut {NoStop}%
\bibitem [{\citenamefont {R{\"{o}}mer}\ \emph {et~al.}(1999)\citenamefont
  {R{\"{o}}mer}, \citenamefont {Schreiber},\ and\ \citenamefont
  {Vojta}}]{roemer1999two}%
  \BibitemOpen
  \bibfield  {author} {\bibinfo {author} {\bibfnamefont {R.A.}\ \bibnamefont
  {R{\"{o}}mer}}, \bibinfo {author} {\bibfnamefont {M.}~\bibnamefont
  {Schreiber}}, \ and\ \bibinfo {author} {\bibfnamefont {T.}~\bibnamefont
  {Vojta}},\ }\bibfield  {title} {\enquote {\bibinfo {title} {Two interacting
  particles in a random potential: Numerical calculations of the interaction
  matrix elements},}\ }\href {\doibase
  10.1002/(SICI)1521-3951(199902)211:2<681::AID-PSSB681>3.0.CO;2-I} {\bibfield
  {journal} {\bibinfo  {journal} {Phys. Stat. Sol. B}\ }\textbf {\bibinfo
  {volume} {211}},\ \bibinfo {pages} {681--691} (\bibinfo {year}
  {1999})}\BibitemShut {NoStop}%
\bibitem [{\citenamefont {Frahm}(1999)}]{frahm1999interaction}%
  \BibitemOpen
  \bibfield  {author} {\bibinfo {author} {\bibfnamefont {K.M.}\ \bibnamefont
  {Frahm}},\ }\bibfield  {title} {\enquote {\bibinfo {title} {Interaction
  induced delocalization of two particles: large system size calculations and
  dependence on interaction strength},}\ }\href {\doibase
  10.1007/s100510050866} {\bibfield  {journal} {\bibinfo  {journal} {Eur. Phys.
  J. B}\ }\textbf {\bibinfo {volume} {10}},\ \bibinfo {pages} {371--378}
  (\bibinfo {year} {1999})}\BibitemShut {NoStop}%
\bibitem [{\citenamefont {Song}\ and\ \citenamefont {von
  Oppen}(1999)}]{song1999general}%
  \BibitemOpen
  \bibfield  {author} {\bibinfo {author} {\bibfnamefont {Pil~Hun}\ \bibnamefont
  {Song}}\ and\ \bibinfo {author} {\bibfnamefont {Felix}\ \bibnamefont {von
  Oppen}},\ }\bibfield  {title} {\enquote {\bibinfo {title} {General
  localization lengths for two interacting particles in a disordered chain},}\
  }\href {\doibase 10.1103/PhysRevB.59.46} {\bibfield  {journal} {\bibinfo
  {journal} {Phys. Rev. B}\ }\textbf {\bibinfo {volume} {59}},\ \bibinfo
  {pages} {46--49} (\bibinfo {year} {1999})}\BibitemShut {NoStop}%
\bibitem [{\citenamefont {Arias}\ \emph {et~al.}(1999)\citenamefont {Arias},
  \citenamefont {Waintal},\ and\ \citenamefont {Pichard}}]{arias1999two}%
  \BibitemOpen
  \bibfield  {author} {\bibinfo {author} {\bibfnamefont {S~De~Toro}\
  \bibnamefont {Arias}}, \bibinfo {author} {\bibfnamefont {Xavier}\
  \bibnamefont {Waintal}}, \ and\ \bibinfo {author} {\bibfnamefont {J-L}\
  \bibnamefont {Pichard}},\ }\bibfield  {title} {\enquote {\bibinfo {title}
  {Two interacting particles in a disordered chain iii: Dynamical aspects of
  the interplay disorder-interaction},}\ }\href
  {https://link.springer.com/article/10.1007/s100510050838} {\bibfield
  {journal} {\bibinfo  {journal} {The European Physical Journal B-Condensed
  Matter and Complex Systems}\ }\textbf {\bibinfo {volume} {10}},\ \bibinfo
  {pages} {149--158} (\bibinfo {year} {1999})}\BibitemShut {NoStop}%
\bibitem [{\citenamefont {Krimer}\ \emph {et~al.}(2011)\citenamefont {Krimer},
  \citenamefont {Khomeriki},\ and\ \citenamefont {Flach}}]{krimer2011two}%
  \BibitemOpen
  \bibfield  {author} {\bibinfo {author} {\bibfnamefont {D.~O.}\ \bibnamefont
  {Krimer}}, \bibinfo {author} {\bibfnamefont {R.}~\bibnamefont {Khomeriki}}, \
  and\ \bibinfo {author} {\bibfnamefont {S.}~\bibnamefont {Flach}},\ }\bibfield
   {title} {\enquote {\bibinfo {title} {Two interacting particles in a random
  potential},}\ }\href {\doibase 10.1134/S0021364011170097} {\bibfield
  {journal} {\bibinfo  {journal} {JETP Lett.}\ }\textbf {\bibinfo {volume}
  {94}},\ \bibinfo {pages} {406--412} (\bibinfo {year} {2011})}\BibitemShut
  {NoStop}%
\bibitem [{\citenamefont {Ivanchenko}\ \emph {et~al.}(2014)\citenamefont
  {Ivanchenko}, \citenamefont {Laptyeva},\ and\ \citenamefont
  {Flach}}]{ivanchenko2014}%
  \BibitemOpen
  \bibfield  {author} {\bibinfo {author} {\bibfnamefont {MV}~\bibnamefont
  {Ivanchenko}}, \bibinfo {author} {\bibfnamefont {TV}~\bibnamefont
  {Laptyeva}}, \ and\ \bibinfo {author} {\bibfnamefont {S}~\bibnamefont
  {Flach}},\ }\bibfield  {title} {\enquote {\bibinfo {title} {Quantum chaotic
  subdiffusion in random potentials},}\ }\href
  {https://journals.aps.org/prb/abstract/10.1103/PhysRevB.89.060301} {\bibfield
   {journal} {\bibinfo  {journal} {Physical Review B}\ }\textbf {\bibinfo
  {volume} {89}},\ \bibinfo {pages} {060301} (\bibinfo {year}
  {2014})}\BibitemShut {NoStop}%
\bibitem [{\citenamefont {Krimer}\ and\ \citenamefont
  {Flach}(2015)}]{krimer2015interaction}%
  \BibitemOpen
  \bibfield  {author} {\bibinfo {author} {\bibfnamefont {D.~O.}\ \bibnamefont
  {Krimer}}\ and\ \bibinfo {author} {\bibfnamefont {S.}~\bibnamefont {Flach}},\
  }\bibfield  {title} {\enquote {\bibinfo {title} {Interaction-induced
  connectivity of disordered two-particle states},}\ }\href {\doibase
  10.1103/PhysRevB.91.100201} {\bibfield  {journal} {\bibinfo  {journal} {Phys.
  Rev. B}\ }\textbf {\bibinfo {volume} {91}},\ \bibinfo {pages} {100201}
  (\bibinfo {year} {2015})}\BibitemShut {NoStop}%
\bibitem [{\citenamefont {Frahm}(2016)}]{frahm2016eigenfunction}%
  \BibitemOpen
  \bibfield  {author} {\bibinfo {author} {\bibfnamefont {Klaus~M.}\
  \bibnamefont {Frahm}},\ }\bibfield  {title} {\enquote {\bibinfo {title}
  {Eigenfunction structure and scaling of two interacting particles in the
  one-dimensional anderson model},}\ }\href {\doibase
  10.1140/epjb/e2016-70114-7} {\bibfield  {journal} {\bibinfo  {journal} {Eur.
  Phys. J. B}\ }\textbf {\bibinfo {volume} {89}},\ \bibinfo {pages} {115}
  (\bibinfo {year} {2016})}\BibitemShut {NoStop}%
\bibitem [{\citenamefont {Yusipov}\ \emph {et~al.}(2017)\citenamefont
  {Yusipov}, \citenamefont {Laptyeva}, \citenamefont {Pirova}, \citenamefont
  {Meyerov}, \citenamefont {Flach},\ and\ \citenamefont
  {Ivanchenko}}]{yusipov2017}%
  \BibitemOpen
  \bibfield  {author} {\bibinfo {author} {\bibfnamefont {Igor~I}\ \bibnamefont
  {Yusipov}}, \bibinfo {author} {\bibfnamefont {Tetyana~V}\ \bibnamefont
  {Laptyeva}}, \bibinfo {author} {\bibfnamefont {Anna~Yu}\ \bibnamefont
  {Pirova}}, \bibinfo {author} {\bibfnamefont {Iosif~B}\ \bibnamefont
  {Meyerov}}, \bibinfo {author} {\bibfnamefont {Sergej}\ \bibnamefont {Flach}},
  \ and\ \bibinfo {author} {\bibfnamefont {Mikhail~V}\ \bibnamefont
  {Ivanchenko}},\ }\bibfield  {title} {\enquote {\bibinfo {title} {Quantum
  subdiffusion with two-and three-body interactions},}\ }\href
  {https://link.springer.com/article/10.1140/epjb/e2017-70722-7} {\bibfield
  {journal} {\bibinfo  {journal} {The European Physical Journal B}\ }\textbf
  {\bibinfo {volume} {90}},\ \bibinfo {pages} {1--7} (\bibinfo {year}
  {2017})}\BibitemShut {NoStop}%
\bibitem [{\citenamefont {Thongjaomayum}\ \emph {et~al.}(2019)\citenamefont
  {Thongjaomayum}, \citenamefont {Andreanov}, \citenamefont {Engl},\ and\
  \citenamefont {Flach}}]{Diana_PhysRevB.100.224203}%
  \BibitemOpen
  \bibfield  {author} {\bibinfo {author} {\bibfnamefont {Diana}\ \bibnamefont
  {Thongjaomayum}}, \bibinfo {author} {\bibfnamefont {Alexei}\ \bibnamefont
  {Andreanov}}, \bibinfo {author} {\bibfnamefont {Thomas}\ \bibnamefont
  {Engl}}, \ and\ \bibinfo {author} {\bibfnamefont {Sergej}\ \bibnamefont
  {Flach}},\ }\bibfield  {title} {\enquote {\bibinfo {title} {Taming two
  interacting particles with disorder},}\ }\href {\doibase
  10.1103/PhysRevB.100.224203} {\bibfield  {journal} {\bibinfo  {journal}
  {Phys. Rev. B}\ }\textbf {\bibinfo {volume} {100}},\ \bibinfo {pages}
  {224203} (\bibinfo {year} {2019})}\BibitemShut {NoStop}%
\bibitem [{\citenamefont {Malishava}\ \emph {et~al.}(2020)\citenamefont
  {Malishava}, \citenamefont {Vakulchyk}, \citenamefont {Fistul},\ and\
  \citenamefont {Flach}}]{PhysRevB.101.144201_merab}%
  \BibitemOpen
  \bibfield  {author} {\bibinfo {author} {\bibfnamefont {Merab}\ \bibnamefont
  {Malishava}}, \bibinfo {author} {\bibfnamefont {Ihor}\ \bibnamefont
  {Vakulchyk}}, \bibinfo {author} {\bibfnamefont {Mikhail}\ \bibnamefont
  {Fistul}}, \ and\ \bibinfo {author} {\bibfnamefont {Sergej}\ \bibnamefont
  {Flach}},\ }\bibfield  {title} {\enquote {\bibinfo {title} {Floquet anderson
  localization of two interacting discrete time quantum walks},}\ }\href
  {\doibase 10.1103/PhysRevB.101.144201} {\bibfield  {journal} {\bibinfo
  {journal} {Phys. Rev. B}\ }\textbf {\bibinfo {volume} {101}},\ \bibinfo
  {pages} {144201} (\bibinfo {year} {2020})}\BibitemShut {NoStop}%
\bibitem [{\citenamefont {Schwiete}\ and\ \citenamefont
  {Finkel'stein}(2013)}]{Schwiete_PhysRevA.88.053611}%
  \BibitemOpen
  \bibfield  {author} {\bibinfo {author} {\bibfnamefont {G.}~\bibnamefont
  {Schwiete}}\ and\ \bibinfo {author} {\bibfnamefont {A.~M.}\ \bibnamefont
  {Finkel'stein}},\ }\bibfield  {title} {\enquote {\bibinfo {title} {Kinetics
  of the disordered bose gas with collisions},}\ }\href {\doibase
  10.1103/PhysRevA.88.053611} {\bibfield  {journal} {\bibinfo  {journal} {Phys.
  Rev. A}\ }\textbf {\bibinfo {volume} {88}},\ \bibinfo {pages} {053611}
  (\bibinfo {year} {2013})}\BibitemShut {NoStop}%
\bibitem [{\citenamefont {Lucioni}\ \emph {et~al.}(2011)\citenamefont
  {Lucioni}, \citenamefont {Deissler}, \citenamefont {Tanzi}, \citenamefont
  {Roati}, \citenamefont {Zaccanti}, \citenamefont {Modugno}, \citenamefont
  {Larcher}, \citenamefont {Dalfovo}, \citenamefont {Inguscio},\ and\
  \citenamefont {Modugno}}]{Lucioni_PhysRevLett.106.230403}%
  \BibitemOpen
  \bibfield  {author} {\bibinfo {author} {\bibfnamefont {E.}~\bibnamefont
  {Lucioni}}, \bibinfo {author} {\bibfnamefont {B.}~\bibnamefont {Deissler}},
  \bibinfo {author} {\bibfnamefont {L.}~\bibnamefont {Tanzi}}, \bibinfo
  {author} {\bibfnamefont {G.}~\bibnamefont {Roati}}, \bibinfo {author}
  {\bibfnamefont {M.}~\bibnamefont {Zaccanti}}, \bibinfo {author}
  {\bibfnamefont {M.}~\bibnamefont {Modugno}}, \bibinfo {author} {\bibfnamefont
  {M.}~\bibnamefont {Larcher}}, \bibinfo {author} {\bibfnamefont
  {F.}~\bibnamefont {Dalfovo}}, \bibinfo {author} {\bibfnamefont
  {M.}~\bibnamefont {Inguscio}}, \ and\ \bibinfo {author} {\bibfnamefont
  {G.}~\bibnamefont {Modugno}},\ }\bibfield  {title} {\enquote {\bibinfo
  {title} {Observation of subdiffusion in a disordered interacting system},}\
  }\href {\doibase 10.1103/PhysRevLett.106.230403} {\bibfield  {journal}
  {\bibinfo  {journal} {Phys. Rev. Lett.}\ }\textbf {\bibinfo {volume} {106}},\
  \bibinfo {pages} {230403} (\bibinfo {year} {2011})}\BibitemShut {NoStop}%
\bibitem [{\citenamefont {Shepelyansky}(1993)}]{PhysRevLett.70.1787}%
  \BibitemOpen
  \bibfield  {author} {\bibinfo {author} {\bibfnamefont {D.~L.}\ \bibnamefont
  {Shepelyansky}},\ }\bibfield  {title} {\enquote {\bibinfo {title}
  {Delocalization of quantum chaos by weak nonlinearity},}\ }\href {\doibase
  10.1103/PhysRevLett.70.1787} {\bibfield  {journal} {\bibinfo  {journal}
  {Phys. Rev. Lett.}\ }\textbf {\bibinfo {volume} {70}},\ \bibinfo {pages}
  {1787--1790} (\bibinfo {year} {1993})}\BibitemShut {NoStop}%
\bibitem [{\citenamefont {Pikovsky}\ and\ \citenamefont
  {Shepelyansky}(2008)}]{PhysRevLett.100.094101}%
  \BibitemOpen
  \bibfield  {author} {\bibinfo {author} {\bibfnamefont {A.~S.}\ \bibnamefont
  {Pikovsky}}\ and\ \bibinfo {author} {\bibfnamefont {D.~L.}\ \bibnamefont
  {Shepelyansky}},\ }\bibfield  {title} {\enquote {\bibinfo {title}
  {Destruction of anderson localization by a weak nonlinearity},}\ }\href
  {\doibase 10.1103/PhysRevLett.100.094101} {\bibfield  {journal} {\bibinfo
  {journal} {Phys. Rev. Lett.}\ }\textbf {\bibinfo {volume} {100}},\ \bibinfo
  {pages} {094101} (\bibinfo {year} {2008})}\BibitemShut {NoStop}%
\bibitem [{\citenamefont {Flach}\ \emph {et~al.}(2009)\citenamefont {Flach},
  \citenamefont {Krimer},\ and\ \citenamefont {Skokos}}]{flach2009universal}%
  \BibitemOpen
  \bibfield  {author} {\bibinfo {author} {\bibfnamefont {Sergej}\ \bibnamefont
  {Flach}}, \bibinfo {author} {\bibfnamefont {DO}~\bibnamefont {Krimer}}, \
  and\ \bibinfo {author} {\bibfnamefont {Ch}~\bibnamefont {Skokos}},\
  }\bibfield  {title} {\enquote {\bibinfo {title} {Universal spreading of wave
  packets in disordered nonlinear systems},}\ }\href
  {https://journals.aps.org/prl/abstract/10.1103/PhysRevLett.102.024101}
  {\bibfield  {journal} {\bibinfo  {journal} {Physical Review Letters}\
  }\textbf {\bibinfo {volume} {102}},\ \bibinfo {pages} {024101} (\bibinfo
  {year} {2009})}\BibitemShut {NoStop}%
\bibitem [{\citenamefont {Laptyeva}\ \emph {et~al.}(2010)\citenamefont
  {Laptyeva}, \citenamefont {Bodyfelt}, \citenamefont {Krimer}, \citenamefont
  {Skokos},\ and\ \citenamefont {Flach}}]{laptyeva2010crossover}%
  \BibitemOpen
  \bibfield  {author} {\bibinfo {author} {\bibfnamefont {T.~V.}\ \bibnamefont
  {Laptyeva}}, \bibinfo {author} {\bibfnamefont {J.~D.}\ \bibnamefont
  {Bodyfelt}}, \bibinfo {author} {\bibfnamefont {D.~O.}\ \bibnamefont
  {Krimer}}, \bibinfo {author} {\bibfnamefont {Ch.}\ \bibnamefont {Skokos}}, \
  and\ \bibinfo {author} {\bibfnamefont {S.}~\bibnamefont {Flach}},\ }\bibfield
   {title} {\enquote {\bibinfo {title} {The crossover from strong to weak chaos
  for nonlinear waves in disordered systems},}\ }\href
  {http://stacks.iop.org/0295-5075/91/i=3/a=30001} {\bibfield  {journal}
  {\bibinfo  {journal} {EPL (Europhysics Letters)}\ }\textbf {\bibinfo {volume}
  {91}},\ \bibinfo {pages} {30001} (\bibinfo {year} {2010})}\BibitemShut
  {NoStop}%
\bibitem [{\citenamefont {Ivanchenko}\ \emph {et~al.}(2011)\citenamefont
  {Ivanchenko}, \citenamefont {Laptyeva},\ and\ \citenamefont
  {Flach}}]{PhysRevLett.107.240602}%
  \BibitemOpen
  \bibfield  {author} {\bibinfo {author} {\bibfnamefont {M.~V.}\ \bibnamefont
  {Ivanchenko}}, \bibinfo {author} {\bibfnamefont {T.~V.}\ \bibnamefont
  {Laptyeva}}, \ and\ \bibinfo {author} {\bibfnamefont {S.}~\bibnamefont
  {Flach}},\ }\bibfield  {title} {\enquote {\bibinfo {title} {Anderson
  localization or nonlinear waves: A matter of probability},}\ }\href {\doibase
  10.1103/PhysRevLett.107.240602} {\bibfield  {journal} {\bibinfo  {journal}
  {Phys. Rev. Lett.}\ }\textbf {\bibinfo {volume} {107}},\ \bibinfo {pages}
  {240602} (\bibinfo {year} {2011})}\BibitemShut {NoStop}%
\bibitem [{\citenamefont {Kati}\ \emph {et~al.}(2020)\citenamefont {Kati},
  \citenamefont {Yu},\ and\ \citenamefont {Flach}}]{kati2020}%
  \BibitemOpen
  \bibfield  {author} {\bibinfo {author} {\bibfnamefont {Yagmur}\ \bibnamefont
  {Kati}}, \bibinfo {author} {\bibfnamefont {Xiaoquan}\ \bibnamefont {Yu}}, \
  and\ \bibinfo {author} {\bibfnamefont {Sergej}\ \bibnamefont {Flach}},\
  }\bibfield  {title} {\enquote {\bibinfo {title} {Density resolved wave packet
  spreading in disordered gross-pitaevskii lattices},}\ }\href
  {https://scipost.org/10.21468/SciPostPhysCore.3.2.006} {\bibfield  {journal}
  {\bibinfo  {journal} {SciPost Phys. Core}\ }\textbf {\bibinfo {volume} {3}},\
  \bibinfo {pages} {006} (\bibinfo {year} {2020})}\BibitemShut {NoStop}%
\bibitem [{\citenamefont {Ermann}\ and\ \citenamefont
  {Shepelyansky}(2021)}]{ErmannChaos2021}%
  \BibitemOpen
  \bibfield  {author} {\bibinfo {author} {\bibfnamefont {Leonardo}\
  \bibnamefont {Ermann}}\ and\ \bibinfo {author} {\bibfnamefont {Dima~L.}\
  \bibnamefont {Shepelyansky}},\ }\bibfield  {title} {\enquote {\bibinfo
  {title} {Deconfinement of classical yang--mills color fields in a disorder
  potential},}\ }\bibfield  {booktitle} {\emph {\bibinfo {booktitle} {Chaos: An
  Interdisciplinary Journal of Nonlinear Science}},\ }\href {\doibase
  10.1063/5.0057969} {\bibfield  {journal} {\bibinfo  {journal} {Chaos: An
  Interdisciplinary Journal of Nonlinear Science}\ }\textbf {\bibinfo {volume}
  {31}},\ \bibinfo {pages} {093106} (\bibinfo {year} {2021})}\BibitemShut
  {NoStop}%
\bibitem [{\citenamefont {Laptyeva}\ \emph {et~al.}(2014)\citenamefont
  {Laptyeva}, \citenamefont {Ivanchenko},\ and\ \citenamefont
  {Flach}}]{Laptyeva_2014}%
  \BibitemOpen
  \bibfield  {author} {\bibinfo {author} {\bibfnamefont {T~V}\ \bibnamefont
  {Laptyeva}}, \bibinfo {author} {\bibfnamefont {M~V}\ \bibnamefont
  {Ivanchenko}}, \ and\ \bibinfo {author} {\bibfnamefont {S}~\bibnamefont
  {Flach}},\ }\bibfield  {title} {\enquote {\bibinfo {title} {Nonlinear lattice
  waves in heterogeneous media},}\ }\href {\doibase
  10.1088/1751-8113/47/49/493001} {\bibfield  {journal} {\bibinfo  {journal}
  {Journal of Physics A: Mathematical and Theoretical}\ }\textbf {\bibinfo
  {volume} {47}},\ \bibinfo {pages} {493001} (\bibinfo {year}
  {2014})}\BibitemShut {NoStop}%
\bibitem [{\citenamefont {Vakulchyk}\ \emph {et~al.}(2017)\citenamefont
  {Vakulchyk}, \citenamefont {Fistul}, \citenamefont {Qin},\ and\ \citenamefont
  {Flach}}]{Ihor_PhysRevB.96.144204_2017}%
  \BibitemOpen
  \bibfield  {author} {\bibinfo {author} {\bibfnamefont {I.}~\bibnamefont
  {Vakulchyk}}, \bibinfo {author} {\bibfnamefont {M.~V.}\ \bibnamefont
  {Fistul}}, \bibinfo {author} {\bibfnamefont {P.}~\bibnamefont {Qin}}, \ and\
  \bibinfo {author} {\bibfnamefont {S.}~\bibnamefont {Flach}},\ }\bibfield
  {title} {\enquote {\bibinfo {title} {Anderson localization in generalized
  discrete-time quantum walks},}\ }\href {\doibase 10.1103/PhysRevB.96.144204}
  {\bibfield  {journal} {\bibinfo  {journal} {Phys. Rev. B}\ }\textbf {\bibinfo
  {volume} {96}},\ \bibinfo {pages} {144204} (\bibinfo {year}
  {2017})}\BibitemShut {NoStop}%
\bibitem [{\citenamefont {Vakulchyk}\ \emph {et~al.}(2019)\citenamefont
  {Vakulchyk}, \citenamefont {Fistul},\ and\ \citenamefont
  {Flach}}]{PhysRevLett.122.040501}%
  \BibitemOpen
  \bibfield  {author} {\bibinfo {author} {\bibfnamefont {Ihor}\ \bibnamefont
  {Vakulchyk}}, \bibinfo {author} {\bibfnamefont {Mikhail~V.}\ \bibnamefont
  {Fistul}}, \ and\ \bibinfo {author} {\bibfnamefont {Sergej}\ \bibnamefont
  {Flach}},\ }\bibfield  {title} {\enquote {\bibinfo {title} {Wave packet
  spreading with disordered nonlinear discrete-time quantum walks},}\ }\href
  {\doibase 10.1103/PhysRevLett.122.040501} {\bibfield  {journal} {\bibinfo
  {journal} {Phys. Rev. Lett.}\ }\textbf {\bibinfo {volume} {122}},\ \bibinfo
  {pages} {040501} (\bibinfo {year} {2019})}\BibitemShut {NoStop}%
\bibitem [{\citenamefont {Gligori{\'c}}\ \emph {et~al.}(2013)\citenamefont
  {Gligori{\'c}}, \citenamefont {Rayanov},\ and\ \citenamefont
  {Flach}}]{gligoric2013make}%
  \BibitemOpen
  \bibfield  {author} {\bibinfo {author} {\bibfnamefont {Goran}\ \bibnamefont
  {Gligori{\'c}}}, \bibinfo {author} {\bibfnamefont {Kristian}\ \bibnamefont
  {Rayanov}}, \ and\ \bibinfo {author} {\bibfnamefont {Sergej}\ \bibnamefont
  {Flach}},\ }\bibfield  {title} {\enquote {\bibinfo {title} {Make slow
  fast---how to speed up interacting disordered matter},}\ }\href
  {https://iopscience.iop.org/article/10.1209/0295-5075/101/10011/meta}
  {\bibfield  {journal} {\bibinfo  {journal} {EPL (Europhysics Letters)}\
  }\textbf {\bibinfo {volume} {101}},\ \bibinfo {pages} {10011} (\bibinfo
  {year} {2013})}\BibitemShut {NoStop}%
\bibitem [{\citenamefont {Chirikov}\ \emph {et~al.}(1981)\citenamefont
  {Chirikov}, \citenamefont {Izrailev},\ and\ \citenamefont
  {Shepelyansky}}]{ChirikovGordon1981}%
  \BibitemOpen
  \bibfield  {author} {\bibinfo {author} {\bibfnamefont {BV}~\bibnamefont
  {Chirikov}}, \bibinfo {author} {\bibfnamefont {FM}~\bibnamefont {Izrailev}},
  \ and\ \bibinfo {author} {\bibfnamefont {DL}~\bibnamefont {Shepelyansky}},\
  }\href {https://www.quantware.ups-tlse.fr/chirikov/refs/chi1981a.pdf}
  {\enquote {\bibinfo {title} {Dynamical stochasticity in classical and quantum
  mechanics soviet scientific reviews c},}\ } (\bibinfo {year}
  {1981})\BibitemShut {NoStop}%
\bibitem [{\citenamefont {Berman}\ \emph {et~al.}(1988)\citenamefont {Berman},
  \citenamefont {Kolovsky},\ and\ \citenamefont
  {Izrailev}}]{BermanPhysicaA1988}%
  \BibitemOpen
  \bibfield  {author} {\bibinfo {author} {\bibfnamefont {GP}~\bibnamefont
  {Berman}}, \bibinfo {author} {\bibfnamefont {AR}~\bibnamefont {Kolovsky}}, \
  and\ \bibinfo {author} {\bibfnamefont {FM}~\bibnamefont {Izrailev}},\
  }\bibfield  {title} {\enquote {\bibinfo {title} {Quantum chaos and
  peculiarities of diffusion in wigner representation},}\ }\href
  {https://www.sciencedirect.com/science/article/pii/0378437188900775}
  {\bibfield  {journal} {\bibinfo  {journal} {Physica A: Statistical Mechanics
  and its Applications}\ }\textbf {\bibinfo {volume} {152}},\ \bibinfo {pages}
  {273--286} (\bibinfo {year} {1988})}\BibitemShut {NoStop}%
\bibitem [{\citenamefont {Berman}\ \emph {et~al.}(1991)\citenamefont {Berman},
  \citenamefont {Kolovski{\u \i}}, \citenamefont {Izrailev},\ and\
  \citenamefont {Iomin}}]{BermanChaos1991}%
  \BibitemOpen
  \bibfield  {author} {\bibinfo {author} {\bibfnamefont {G.~P.}\ \bibnamefont
  {Berman}}, \bibinfo {author} {\bibfnamefont {A.~R.}\ \bibnamefont
  {Kolovski{\u \i}}}, \bibinfo {author} {\bibfnamefont {F.~M.}\ \bibnamefont
  {Izrailev}}, \ and\ \bibinfo {author} {\bibfnamefont {A.~M.}\ \bibnamefont
  {Iomin}},\ }\bibfield  {title} {\enquote {\bibinfo {title} {Quantum chaos in
  the wigner representation},}\ }\bibfield  {booktitle} {\emph {\bibinfo
  {booktitle} {Chaos: An Interdisciplinary Journal of Nonlinear Science}},\
  }\href {\doibase 10.1063/1.165832} {\bibfield  {journal} {\bibinfo  {journal}
  {Chaos: An Interdisciplinary Journal of Nonlinear Science}\ }\textbf
  {\bibinfo {volume} {1}},\ \bibinfo {pages} {220--223} (\bibinfo {year}
  {1991})}\BibitemShut {NoStop}%
\bibitem [{\citenamefont {Guarneri}\ \emph {et~al.}(2014)\citenamefont
  {Guarneri}, \citenamefont {Casati},\ and\ \citenamefont
  {Karle}}]{GuarneriPRL2014}%
  \BibitemOpen
  \bibfield  {author} {\bibinfo {author} {\bibfnamefont {Italo}\ \bibnamefont
  {Guarneri}}, \bibinfo {author} {\bibfnamefont {Giulio}\ \bibnamefont
  {Casati}}, \ and\ \bibinfo {author} {\bibfnamefont {Volker}\ \bibnamefont
  {Karle}},\ }\bibfield  {title} {\enquote {\bibinfo {title} {Classical
  dynamical localization},}\ }\href
  {https://journals.aps.org/prl/abstract/10.1103/PhysRevLett.113.174101}
  {\bibfield  {journal} {\bibinfo  {journal} {Physical review letters}\
  }\textbf {\bibinfo {volume} {113}},\ \bibinfo {pages} {174101} (\bibinfo
  {year} {2014})}\BibitemShut {NoStop}%
\bibitem [{\citenamefont {Aharonov}\ \emph {et~al.}(1993)\citenamefont
  {Aharonov}, \citenamefont {Davidovich},\ and\ \citenamefont
  {Zagury}}]{Aharonov_PhysRevA.48.1687}%
  \BibitemOpen
  \bibfield  {author} {\bibinfo {author} {\bibfnamefont {Y.}~\bibnamefont
  {Aharonov}}, \bibinfo {author} {\bibfnamefont {L.}~\bibnamefont
  {Davidovich}}, \ and\ \bibinfo {author} {\bibfnamefont {N.}~\bibnamefont
  {Zagury}},\ }\bibfield  {title} {\enquote {\bibinfo {title} {Quantum random
  walks},}\ }\href {\doibase 10.1103/PhysRevA.48.1687} {\bibfield  {journal}
  {\bibinfo  {journal} {Phys. Rev. A}\ }\textbf {\bibinfo {volume} {48}},\
  \bibinfo {pages} {1687--1690} (\bibinfo {year} {1993})}\BibitemShut {NoStop}%
\bibitem [{\citenamefont {Meyer}(1996)}]{Meyer_1996}%
  \BibitemOpen
  \bibfield  {author} {\bibinfo {author} {\bibfnamefont {David~A.}\
  \bibnamefont {Meyer}},\ }\bibfield  {title} {\enquote {\bibinfo {title} {From
  quantum cellular automata to quantum lattice gases},}\ }\href {\doibase
  10.1007/bf02199356} {\bibfield  {journal} {\bibinfo  {journal} {Journal of
  Statistical Physics}\ }\textbf {\bibinfo {volume} {85}},\ \bibinfo {pages}
  {551--574} (\bibinfo {year} {1996})}\BibitemShut {NoStop}%
\bibitem [{\citenamefont {Mallick}\ and\ \citenamefont
  {Chandrashekar}(2016)}]{Mallick_2016}%
  \BibitemOpen
  \bibfield  {author} {\bibinfo {author} {\bibfnamefont {Arindam}\ \bibnamefont
  {Mallick}}\ and\ \bibinfo {author} {\bibfnamefont {C.~M.}\ \bibnamefont
  {Chandrashekar}},\ }\bibfield  {title} {\enquote {\bibinfo {title} {Dirac
  cellular automaton from split-step quantum walk},}\ }\href {\doibase
  10.1038/srep25779} {\bibfield  {journal} {\bibinfo  {journal} {Scientific
  Reports}\ }\textbf {\bibinfo {volume} {6}} (\bibinfo {year} {2016}),\
  10.1038/srep25779}\BibitemShut {NoStop}%
\bibitem [{\citenamefont {Mallick}(2019)}]{mallick2019quantum}%
  \BibitemOpen
  \bibfield  {author} {\bibinfo {author} {\bibfnamefont {Arindam}\ \bibnamefont
  {Mallick}},\ }\href@noop {} {\enquote {\bibinfo {title} {Quantum simulation
  of neutrino oscillation and dirac particle dynamics in curved space-time},}\
  } (\bibinfo {year} {2019}),\ \Eprint {http://arxiv.org/abs/1901.04014}
  {arXiv:1901.04014 [quant-ph]} \BibitemShut {NoStop}%
\bibitem [{\citenamefont {Arnault}\ and\ \citenamefont
  {Debbasch}(2016)}]{Arnault_2016}%
  \BibitemOpen
  \bibfield  {author} {\bibinfo {author} {\bibfnamefont {Pablo}\ \bibnamefont
  {Arnault}}\ and\ \bibinfo {author} {\bibfnamefont {Fabrice}\ \bibnamefont
  {Debbasch}},\ }\bibfield  {title} {\enquote {\bibinfo {title} {Quantum walks
  and discrete gauge theories},}\ }\href {\doibase 10.1103/physreva.93.052301}
  {\bibfield  {journal} {\bibinfo  {journal} {Physical Review A}\ }\textbf
  {\bibinfo {volume} {93}} (\bibinfo {year} {2016}),\
  10.1103/physreva.93.052301}\BibitemShut {NoStop}%
\bibitem [{\citenamefont {Kitagawa}\ \emph {et~al.}(2010)\citenamefont
  {Kitagawa}, \citenamefont {Rudner}, \citenamefont {Berg},\ and\ \citenamefont
  {Demler}}]{Kitagawa_2010}%
  \BibitemOpen
  \bibfield  {author} {\bibinfo {author} {\bibfnamefont {Takuya}\ \bibnamefont
  {Kitagawa}}, \bibinfo {author} {\bibfnamefont {Mark~S.}\ \bibnamefont
  {Rudner}}, \bibinfo {author} {\bibfnamefont {Erez}\ \bibnamefont {Berg}}, \
  and\ \bibinfo {author} {\bibfnamefont {Eugene}\ \bibnamefont {Demler}},\
  }\bibfield  {title} {\enquote {\bibinfo {title} {Exploring topological phases
  with quantum walks},}\ }\href {\doibase 10.1103/physreva.82.033429}
  {\bibfield  {journal} {\bibinfo  {journal} {Physical Review A}\ }\textbf
  {\bibinfo {volume} {82}} (\bibinfo {year} {2010}),\
  10.1103/physreva.82.033429}\BibitemShut {NoStop}%
\bibitem [{\citenamefont {Asb\'oth}(2012)}]{Asboth_PhysRevB.86.195414}%
  \BibitemOpen
  \bibfield  {author} {\bibinfo {author} {\bibfnamefont {J.~K.}\ \bibnamefont
  {Asb\'oth}},\ }\bibfield  {title} {\enquote {\bibinfo {title} {Symmetries,
  topological phases, and bound states in the one-dimensional quantum walk},}\
  }\href {\doibase 10.1103/PhysRevB.86.195414} {\bibfield  {journal} {\bibinfo
  {journal} {Phys. Rev. B}\ }\textbf {\bibinfo {volume} {86}},\ \bibinfo
  {pages} {195414} (\bibinfo {year} {2012})}\BibitemShut {NoStop}%
\bibitem [{\citenamefont {Asb\'oth}\ and\ \citenamefont
  {Mallick}(2020)}]{Asboth_PhysRevB.102.224202}%
  \BibitemOpen
  \bibfield  {author} {\bibinfo {author} {\bibfnamefont {J\'anos~K.}\
  \bibnamefont {Asb\'oth}}\ and\ \bibinfo {author} {\bibfnamefont {Arindam}\
  \bibnamefont {Mallick}},\ }\bibfield  {title} {\enquote {\bibinfo {title}
  {Topological delocalization in the completely disordered two-dimensional
  quantum walk},}\ }\href {\doibase 10.1103/PhysRevB.102.224202} {\bibfield
  {journal} {\bibinfo  {journal} {Phys. Rev. B}\ }\textbf {\bibinfo {volume}
  {102}},\ \bibinfo {pages} {224202} (\bibinfo {year} {2020})}\BibitemShut
  {NoStop}%
\bibitem [{\citenamefont {Chandrashekar}\ and\ \citenamefont
  {Busch}(2014)}]{chandrashekar2014quantum}%
  \BibitemOpen
  \bibfield  {author} {\bibinfo {author} {\bibfnamefont {CM}~\bibnamefont
  {Chandrashekar}}\ and\ \bibinfo {author} {\bibfnamefont {Th}~\bibnamefont
  {Busch}},\ }\bibfield  {title} {\enquote {\bibinfo {title} {Quantum
  percolation and transition point of a directed discrete-time quantum walk},}\
  }\href {https://www.nature.com/articles/srep06583} {\bibfield  {journal}
  {\bibinfo  {journal} {Scientific reports}\ }\textbf {\bibinfo {volume} {4}},\
  \bibinfo {pages} {1--11} (\bibinfo {year} {2014})}\BibitemShut {NoStop}%
\bibitem [{\citenamefont {Koll{\'a}r}\ \emph {et~al.}(2014)\citenamefont
  {Koll{\'a}r}, \citenamefont {Novotn{\'y}}, \citenamefont {Kiss},\ and\
  \citenamefont {Jex}}]{Koll_r_2014}%
  \BibitemOpen
  \bibfield  {author} {\bibinfo {author} {\bibfnamefont {B{\'a}lint}\
  \bibnamefont {Koll{\'a}r}}, \bibinfo {author} {\bibfnamefont {Jaroslav}\
  \bibnamefont {Novotn{\'y}}}, \bibinfo {author} {\bibfnamefont {Tam{\'a}s}\
  \bibnamefont {Kiss}}, \ and\ \bibinfo {author} {\bibfnamefont {Igor}\
  \bibnamefont {Jex}},\ }\bibfield  {title} {\enquote {\bibinfo {title}
  {Discrete time quantum walks on percolation graphs},}\ }\href {\doibase
  10.1140/epjp/i2014-14103-6} {\bibfield  {journal} {\bibinfo  {journal} {The
  European Physical Journal Plus}\ }\textbf {\bibinfo {volume} {129}} (\bibinfo
  {year} {2014}),\ 10.1140/epjp/i2014-14103-6}\BibitemShut {NoStop}%
\bibitem [{\citenamefont {Koll{\'a}r}\ \emph {et~al.}(2020)\citenamefont
  {Koll{\'a}r}, \citenamefont {Gily{\'e}n}, \citenamefont {Tk{\'a}{\v
  c}ov{\'a}}, \citenamefont {Kiss}, \citenamefont {Jex},\ and\ \citenamefont
  {{\v S}tefa{\v n}{\'a}k}}]{Koll_r_2020}%
  \BibitemOpen
  \bibfield  {author} {\bibinfo {author} {\bibfnamefont {B.}~\bibnamefont
  {Koll{\'a}r}}, \bibinfo {author} {\bibfnamefont {A.}~\bibnamefont
  {Gily{\'e}n}}, \bibinfo {author} {\bibfnamefont {I.}~\bibnamefont {Tk{\'a}{\v
  c}ov{\'a}}}, \bibinfo {author} {\bibfnamefont {T.}~\bibnamefont {Kiss}},
  \bibinfo {author} {\bibfnamefont {I.}~\bibnamefont {Jex}}, \ and\ \bibinfo
  {author} {\bibfnamefont {M.}~\bibnamefont {{\v S}tefa{\v n}{\'a}k}},\
  }\bibfield  {title} {\enquote {\bibinfo {title} {Complete classification of
  trapping coins for quantum walks on the two-dimensional square lattice},}\
  }\href {\doibase 10.1103/physreva.102.012207} {\bibfield  {journal} {\bibinfo
   {journal} {Physical Review A}\ }\textbf {\bibinfo {volume} {102}} (\bibinfo
  {year} {2020}),\ 10.1103/physreva.102.012207}\BibitemShut {NoStop}%
\bibitem [{\citenamefont {Chawla}\ \emph {et~al.}(2020)\citenamefont {Chawla},
  \citenamefont {Mangal},\ and\ \citenamefont
  {Chandrashekar}}]{chawla2020discrete}%
  \BibitemOpen
  \bibfield  {author} {\bibinfo {author} {\bibfnamefont {Prateek}\ \bibnamefont
  {Chawla}}, \bibinfo {author} {\bibfnamefont {Roopesh}\ \bibnamefont
  {Mangal}}, \ and\ \bibinfo {author} {\bibfnamefont {CM}~\bibnamefont
  {Chandrashekar}},\ }\bibfield  {title} {\enquote {\bibinfo {title}
  {Discrete-time quantum walk algorithm for ranking nodes on a network},}\
  }\href {https://link.springer.com/article/10.1007%2Fs11128-020-02650-4}
  {\bibfield  {journal} {\bibinfo  {journal} {Quantum Information Processing}\
  }\textbf {\bibinfo {volume} {19}},\ \bibinfo {pages} {1--21} (\bibinfo {year}
  {2020})}\BibitemShut {NoStop}%
\bibitem [{\citenamefont {Srikara}\ and\ \citenamefont
  {Chandrashekar}(2020)}]{Srikara_2020}%
  \BibitemOpen
  \bibfield  {author} {\bibinfo {author} {\bibfnamefont {S.}~\bibnamefont
  {Srikara}}\ and\ \bibinfo {author} {\bibfnamefont {C.~M.}\ \bibnamefont
  {Chandrashekar}},\ }\bibfield  {title} {\enquote {\bibinfo {title} {Quantum
  direct communication protocols using discrete-time quantum walk},}\ }\href
  {\doibase 10.1007/s11128-020-02793-4} {\bibfield  {journal} {\bibinfo
  {journal} {Quantum Information Processing}\ }\textbf {\bibinfo {volume} {19}}
  (\bibinfo {year} {2020}),\ 10.1007/s11128-020-02793-4}\BibitemShut {NoStop}%
\bibitem [{\citenamefont {Vlachou}\ \emph {et~al.}(2018)\citenamefont
  {Vlachou}, \citenamefont {Krawec}, \citenamefont {Mateus}, \citenamefont
  {Paunkovi{\'c}},\ and\ \citenamefont {Souto}}]{Vlachou_2018}%
  \BibitemOpen
  \bibfield  {author} {\bibinfo {author} {\bibfnamefont {Chrysoula}\
  \bibnamefont {Vlachou}}, \bibinfo {author} {\bibfnamefont {Walter}\
  \bibnamefont {Krawec}}, \bibinfo {author} {\bibfnamefont {Paulo}\
  \bibnamefont {Mateus}}, \bibinfo {author} {\bibfnamefont {Nikola}\
  \bibnamefont {Paunkovi{\'c}}}, \ and\ \bibinfo {author} {\bibfnamefont
  {Andr{\'e}}\ \bibnamefont {Souto}},\ }\bibfield  {title} {\enquote {\bibinfo
  {title} {Quantum key distribution with quantum walks},}\ }\href {\doibase
  10.1007/s11128-018-2055-y} {\bibfield  {journal} {\bibinfo  {journal}
  {Quantum Information Processing}\ }\textbf {\bibinfo {volume} {17}} (\bibinfo
  {year} {2018}),\ 10.1007/s11128-018-2055-y}\BibitemShut {NoStop}%
\bibitem [{\citenamefont {Ryan}\ \emph {et~al.}(2005)\citenamefont {Ryan},
  \citenamefont {Laforest}, \citenamefont {Boileau},\ and\ \citenamefont
  {Laflamme}}]{Ryan_2005}%
  \BibitemOpen
  \bibfield  {author} {\bibinfo {author} {\bibfnamefont {C.~A.}\ \bibnamefont
  {Ryan}}, \bibinfo {author} {\bibfnamefont {M.}~\bibnamefont {Laforest}},
  \bibinfo {author} {\bibfnamefont {J.~C.}\ \bibnamefont {Boileau}}, \ and\
  \bibinfo {author} {\bibfnamefont {R.}~\bibnamefont {Laflamme}},\ }\bibfield
  {title} {\enquote {\bibinfo {title} {Experimental implementation of a
  discrete-time quantum random walk on an nmr quantum-information processor},}\
  }\href {\doibase 10.1103/physreva.72.062317} {\bibfield  {journal} {\bibinfo
  {journal} {Physical Review A}\ }\textbf {\bibinfo {volume} {72}} (\bibinfo
  {year} {2005}),\ 10.1103/physreva.72.062317}\BibitemShut {NoStop}%
\bibitem [{\citenamefont {Crespi}\ \emph {et~al.}(2013)\citenamefont {Crespi},
  \citenamefont {Osellame}, \citenamefont {Ramponi}, \citenamefont
  {Giovannetti}, \citenamefont {Fazio}, \citenamefont {Sansoni}, \citenamefont
  {De~Nicola}, \citenamefont {Sciarrino},\ and\ \citenamefont
  {Mataloni}}]{Crespi_2013}%
  \BibitemOpen
  \bibfield  {author} {\bibinfo {author} {\bibfnamefont {Andrea}\ \bibnamefont
  {Crespi}}, \bibinfo {author} {\bibfnamefont {Roberto}\ \bibnamefont
  {Osellame}}, \bibinfo {author} {\bibfnamefont {Roberta}\ \bibnamefont
  {Ramponi}}, \bibinfo {author} {\bibfnamefont {Vittorio}\ \bibnamefont
  {Giovannetti}}, \bibinfo {author} {\bibfnamefont {Rosario}\ \bibnamefont
  {Fazio}}, \bibinfo {author} {\bibfnamefont {Linda}\ \bibnamefont {Sansoni}},
  \bibinfo {author} {\bibfnamefont {Francesco}\ \bibnamefont {De~Nicola}},
  \bibinfo {author} {\bibfnamefont {Fabio}\ \bibnamefont {Sciarrino}}, \ and\
  \bibinfo {author} {\bibfnamefont {Paolo}\ \bibnamefont {Mataloni}},\
  }\bibfield  {title} {\enquote {\bibinfo {title} {Anderson localization of
  entangled photons in an integrated quantum walk},}\ }\href {\doibase
  10.1038/nphoton.2013.26} {\bibfield  {journal} {\bibinfo  {journal} {Nature
  Photonics}\ }\textbf {\bibinfo {volume} {7}},\ \bibinfo {pages} {322--328}
  (\bibinfo {year} {2013})}\BibitemShut {NoStop}%
\bibitem [{\citenamefont {Su}\ \emph {et~al.}(2019)\citenamefont {Su},
  \citenamefont {Zhang}, \citenamefont {Yu}, \citenamefont {Zhou},
  \citenamefont {Jin}, \citenamefont {Xu}, \citenamefont {Xiong}, \citenamefont
  {Xu}, \citenamefont {Sun}, \citenamefont {Chen},\ and\ \citenamefont
  {et~al.}}]{Su_2019}%
  \BibitemOpen
  \bibfield  {author} {\bibinfo {author} {\bibfnamefont {Qi-Ping}\ \bibnamefont
  {Su}}, \bibinfo {author} {\bibfnamefont {Yu}~\bibnamefont {Zhang}}, \bibinfo
  {author} {\bibfnamefont {Li}~\bibnamefont {Yu}}, \bibinfo {author}
  {\bibfnamefont {Jia-Qi}\ \bibnamefont {Zhou}}, \bibinfo {author}
  {\bibfnamefont {Jin-Shuang}\ \bibnamefont {Jin}}, \bibinfo {author}
  {\bibfnamefont {Xiao-Qiang}\ \bibnamefont {Xu}}, \bibinfo {author}
  {\bibfnamefont {Shao-Jie}\ \bibnamefont {Xiong}}, \bibinfo {author}
  {\bibfnamefont {QingJun}\ \bibnamefont {Xu}}, \bibinfo {author}
  {\bibfnamefont {Zhe}\ \bibnamefont {Sun}}, \bibinfo {author} {\bibfnamefont
  {Kefei}\ \bibnamefont {Chen}}, \ and\ \bibinfo {author} {\bibnamefont
  {et~al.}},\ }\bibfield  {title} {\enquote {\bibinfo {title} {Experimental
  demonstration of quantum walks with initial superposition states},}\ }\href
  {\doibase 10.1038/s41534-019-0155-x} {\bibfield  {journal} {\bibinfo
  {journal} {npj Quantum Information}\ }\textbf {\bibinfo {volume} {5}}
  (\bibinfo {year} {2019}),\ 10.1038/s41534-019-0155-x}\BibitemShut {NoStop}%
\bibitem [{\citenamefont {Acasiete}\ \emph {et~al.}(2020)\citenamefont
  {Acasiete}, \citenamefont {Agostini}, \citenamefont {Moqadam},\ and\
  \citenamefont {Portugal}}]{Acasiete_2020}%
  \BibitemOpen
  \bibfield  {author} {\bibinfo {author} {\bibfnamefont {F.}~\bibnamefont
  {Acasiete}}, \bibinfo {author} {\bibfnamefont {F.~P.}\ \bibnamefont
  {Agostini}}, \bibinfo {author} {\bibfnamefont {J.~Khatibi}\ \bibnamefont
  {Moqadam}}, \ and\ \bibinfo {author} {\bibfnamefont {R.}~\bibnamefont
  {Portugal}},\ }\bibfield  {title} {\enquote {\bibinfo {title} {Implementation
  of quantum walks on ibm quantum computers},}\ }\href {\doibase
  10.1007/s11128-020-02938-5} {\bibfield  {journal} {\bibinfo  {journal}
  {Quantum Information Processing}\ }\textbf {\bibinfo {volume} {19}} (\bibinfo
  {year} {2020}),\ 10.1007/s11128-020-02938-5}\BibitemShut {NoStop}%
\bibitem [{\citenamefont {Huerta~Alderete}\ \emph {et~al.}(2020)\citenamefont
  {Huerta~Alderete}, \citenamefont {Singh}, \citenamefont {Nguyen},
  \citenamefont {Zhu}, \citenamefont {Balu}, \citenamefont {Monroe},
  \citenamefont {Chandrashekar},\ and\ \citenamefont
  {Linke}}]{Huerta_Alderete_2020}%
  \BibitemOpen
  \bibfield  {author} {\bibinfo {author} {\bibfnamefont {C.}~\bibnamefont
  {Huerta~Alderete}}, \bibinfo {author} {\bibfnamefont {Shivani}\ \bibnamefont
  {Singh}}, \bibinfo {author} {\bibfnamefont {Nhung~H.}\ \bibnamefont
  {Nguyen}}, \bibinfo {author} {\bibfnamefont {Daiwei}\ \bibnamefont {Zhu}},
  \bibinfo {author} {\bibfnamefont {Radhakrishnan}\ \bibnamefont {Balu}},
  \bibinfo {author} {\bibfnamefont {Christopher}\ \bibnamefont {Monroe}},
  \bibinfo {author} {\bibfnamefont {C.~M.}\ \bibnamefont {Chandrashekar}}, \
  and\ \bibinfo {author} {\bibfnamefont {Norbert~M.}\ \bibnamefont {Linke}},\
  }\bibfield  {title} {\enquote {\bibinfo {title} {Quantum walks and dirac
  cellular automata on a programmable trapped-ion quantum computer},}\ }\href
  {\doibase 10.1038/s41467-020-17519-4} {\bibfield  {journal} {\bibinfo
  {journal} {Nature Communications}\ }\textbf {\bibinfo {volume} {11}}
  (\bibinfo {year} {2020}),\ 10.1038/s41467-020-17519-4}\BibitemShut {NoStop}%
\bibitem [{\citenamefont {Navarrete-Benlloch}\ \emph
  {et~al.}(2007)\citenamefont {Navarrete-Benlloch}, \citenamefont {P\'erez},\
  and\ \citenamefont {Rold\'an}}]{Navarrete_PhysRevA.75.062333}%
  \BibitemOpen
  \bibfield  {author} {\bibinfo {author} {\bibfnamefont {C.}~\bibnamefont
  {Navarrete-Benlloch}}, \bibinfo {author} {\bibfnamefont {A.}~\bibnamefont
  {P\'erez}}, \ and\ \bibinfo {author} {\bibfnamefont {Eugenio}\ \bibnamefont
  {Rold\'an}},\ }\bibfield  {title} {\enquote {\bibinfo {title} {Nonlinear
  optical galton board},}\ }\href {\doibase 10.1103/PhysRevA.75.062333}
  {\bibfield  {journal} {\bibinfo  {journal} {Phys. Rev. A}\ }\textbf {\bibinfo
  {volume} {75}},\ \bibinfo {pages} {062333} (\bibinfo {year}
  {2007})}\BibitemShut {NoStop}%
\bibitem [{\citenamefont {Maeda}\ \emph {et~al.}(2019)\citenamefont {Maeda},
  \citenamefont {Sasaki}, \citenamefont {Segawa}, \citenamefont {Suzuki},\ and\
  \citenamefont {Suzuki}}]{Maeda_2019}%
  \BibitemOpen
  \bibfield  {author} {\bibinfo {author} {\bibfnamefont {Masaya}\ \bibnamefont
  {Maeda}}, \bibinfo {author} {\bibfnamefont {Hironobu}\ \bibnamefont
  {Sasaki}}, \bibinfo {author} {\bibfnamefont {Etsuo}\ \bibnamefont {Segawa}},
  \bibinfo {author} {\bibfnamefont {Akito}\ \bibnamefont {Suzuki}}, \ and\
  \bibinfo {author} {\bibfnamefont {Kanako}\ \bibnamefont {Suzuki}},\
  }\bibfield  {title} {\enquote {\bibinfo {title} {Dynamics of solitons for
  nonlinear quantum walks},}\ }\href {\doibase 10.1088/2399-6528/aafe2c}
  {\bibfield  {journal} {\bibinfo  {journal} {Journal of Physics
  Communications}\ }\textbf {\bibinfo {volume} {3}},\ \bibinfo {pages} {075002}
  (\bibinfo {year} {2019})}\BibitemShut {NoStop}%
\bibitem [{\citenamefont {Maeda}\ and\ \citenamefont
  {Suzuki}(2020)}]{maeda2020continuous}%
  \BibitemOpen
  \bibfield  {author} {\bibinfo {author} {\bibfnamefont {Masaya}\ \bibnamefont
  {Maeda}}\ and\ \bibinfo {author} {\bibfnamefont {Akito}\ \bibnamefont
  {Suzuki}},\ }\bibfield  {title} {\enquote {\bibinfo {title} {Continuous
  limits of linear and nonlinear quantum walks},}\ }\href
  {https://www.worldscientific.com/doi/10.1142/S0129055X20500087} {\bibfield
  {journal} {\bibinfo  {journal} {Reviews in Mathematical Physics}\ }\textbf
  {\bibinfo {volume} {32}},\ \bibinfo {pages} {2050008} (\bibinfo {year}
  {2020})}\BibitemShut {NoStop}%
\bibitem [{\citenamefont {Mochizuki}\ \emph {et~al.}(2020)\citenamefont
  {Mochizuki}, \citenamefont {Kawakami},\ and\ \citenamefont
  {Obuse}}]{mochizuki2020stability}%
  \BibitemOpen
  \bibfield  {author} {\bibinfo {author} {\bibfnamefont {Ken}\ \bibnamefont
  {Mochizuki}}, \bibinfo {author} {\bibfnamefont {Norio}\ \bibnamefont
  {Kawakami}}, \ and\ \bibinfo {author} {\bibfnamefont {Hideaki}\ \bibnamefont
  {Obuse}},\ }\bibfield  {title} {\enquote {\bibinfo {title} {Stability of
  topologically protected edge states in nonlinear quantum walks: additional
  bifurcations unique to floquet systems},}\ }\href
  {https://iopscience.iop.org/article/10.1088/1751-8121/ab6514} {\bibfield
  {journal} {\bibinfo  {journal} {Journal of Physics A: Mathematical and
  Theoretical}\ }\textbf {\bibinfo {volume} {53}},\ \bibinfo {pages} {085702}
  (\bibinfo {year} {2020})}\BibitemShut {NoStop}%
\bibitem [{\citenamefont {Adami}\ \emph {et~al.}(2019)\citenamefont {Adami},
  \citenamefont {Fukuizumi},\ and\ \citenamefont
  {Segawa}}]{adami2019nonlinear}%
  \BibitemOpen
  \bibfield  {author} {\bibinfo {author} {\bibfnamefont {Riccardo}\
  \bibnamefont {Adami}}, \bibinfo {author} {\bibfnamefont {Reika}\ \bibnamefont
  {Fukuizumi}}, \ and\ \bibinfo {author} {\bibfnamefont {Etsuo}\ \bibnamefont
  {Segawa}},\ }\bibfield  {title} {\enquote {\bibinfo {title} {A nonlinear
  quantum walk induced by a quantum graph with nonlinear delta potentials},}\
  }\href {https://link.springer.com/article/10.1007/s11128-019-2215-8}
  {\bibfield  {journal} {\bibinfo  {journal} {Quantum Information Processing}\
  }\textbf {\bibinfo {volume} {18}},\ \bibinfo {pages} {119} (\bibinfo {year}
  {2019})}\BibitemShut {NoStop}%
\bibitem [{\citenamefont {Flach}(2010)}]{FLACH2010548}%
  \BibitemOpen
  \bibfield  {author} {\bibinfo {author} {\bibfnamefont {S.}~\bibnamefont
  {Flach}},\ }\bibfield  {title} {\enquote {\bibinfo {title} {Spreading of
  waves in nonlinear disordered media},}\ }\href {\doibase
  https://doi.org/10.1016/j.chemphys.2010.02.022} {\bibfield  {journal}
  {\bibinfo  {journal} {Chemical Physics}\ }\textbf {\bibinfo {volume} {375}},\
  \bibinfo {pages} {548 -- 556} (\bibinfo {year} {2010})},\ \bibinfo {note}
  {stochastic processes in Physics and Chemistry (in honor of Peter
  H{\"a}nggi)}\BibitemShut {NoStop}%
\bibitem [{\citenamefont {Inouye}\ \emph {et~al.}(1998)\citenamefont {Inouye},
  \citenamefont {Andrews}, \citenamefont {Stenger}, \citenamefont {Miesner},
  \citenamefont {Stamper-Kurn},\ and\ \citenamefont
  {Ketterle}}]{inouye1998observation}%
  \BibitemOpen
  \bibfield  {author} {\bibinfo {author} {\bibfnamefont {S}~\bibnamefont
  {Inouye}}, \bibinfo {author} {\bibfnamefont {MR}~\bibnamefont {Andrews}},
  \bibinfo {author} {\bibfnamefont {J}~\bibnamefont {Stenger}}, \bibinfo
  {author} {\bibfnamefont {H-J}\ \bibnamefont {Miesner}}, \bibinfo {author}
  {\bibfnamefont {Dan~M}\ \bibnamefont {Stamper-Kurn}}, \ and\ \bibinfo
  {author} {\bibfnamefont {W}~\bibnamefont {Ketterle}},\ }\bibfield  {title}
  {\enquote {\bibinfo {title} {Observation of feshbach resonances in a
  bose--einstein condensate},}\ }\href {https://www.nature.com/articles/32354}
  {\bibfield  {journal} {\bibinfo  {journal} {Nature}\ }\textbf {\bibinfo
  {volume} {392}},\ \bibinfo {pages} {151--154} (\bibinfo {year}
  {1998})}\BibitemShut {NoStop}%
\bibitem [{\citenamefont {Khaykovich}\ \emph {et~al.}(2002)\citenamefont
  {Khaykovich}, \citenamefont {Schreck}, \citenamefont {Ferrari}, \citenamefont
  {Bourdel}, \citenamefont {Cubizolles}, \citenamefont {Carr}, \citenamefont
  {Castin},\ and\ \citenamefont {Salomon}}]{khaykovich2002formation}%
  \BibitemOpen
  \bibfield  {author} {\bibinfo {author} {\bibfnamefont {L}~\bibnamefont
  {Khaykovich}}, \bibinfo {author} {\bibfnamefont {F}~\bibnamefont {Schreck}},
  \bibinfo {author} {\bibfnamefont {G}~\bibnamefont {Ferrari}}, \bibinfo
  {author} {\bibfnamefont {Thomas}\ \bibnamefont {Bourdel}}, \bibinfo {author}
  {\bibfnamefont {Julien}\ \bibnamefont {Cubizolles}}, \bibinfo {author}
  {\bibfnamefont {Lincoln~D}\ \bibnamefont {Carr}}, \bibinfo {author}
  {\bibfnamefont {Yvan}\ \bibnamefont {Castin}}, \ and\ \bibinfo {author}
  {\bibfnamefont {Christophe}\ \bibnamefont {Salomon}},\ }\bibfield  {title}
  {\enquote {\bibinfo {title} {Formation of a matter-wave bright soliton},}\
  }\href {https://www.science.org/doi/10.1126/science.1071021} {\bibfield
  {journal} {\bibinfo  {journal} {Science}\ }\textbf {\bibinfo {volume}
  {296}},\ \bibinfo {pages} {1290--1293} (\bibinfo {year} {2002})}\BibitemShut
  {NoStop}%
\bibitem [{\citenamefont {Weber}\ \emph {et~al.}(2003)\citenamefont {Weber},
  \citenamefont {Herbig}, \citenamefont {Mark}, \citenamefont {N{\"a}gerl},\
  and\ \citenamefont {Grimm}}]{weber2003bose}%
  \BibitemOpen
  \bibfield  {author} {\bibinfo {author} {\bibfnamefont {Tino}\ \bibnamefont
  {Weber}}, \bibinfo {author} {\bibfnamefont {Jens}\ \bibnamefont {Herbig}},
  \bibinfo {author} {\bibfnamefont {Michael}\ \bibnamefont {Mark}}, \bibinfo
  {author} {\bibfnamefont {Hanns-Christoph}\ \bibnamefont {N{\"a}gerl}}, \ and\
  \bibinfo {author} {\bibfnamefont {Rudolf}\ \bibnamefont {Grimm}},\ }\bibfield
   {title} {\enquote {\bibinfo {title} {Bose-einstein condensation of
  cesium},}\ }\href {https://www.science.org/doi/10.1126/science.1079699}
  {\bibfield  {journal} {\bibinfo  {journal} {Science}\ }\textbf {\bibinfo
  {volume} {299}},\ \bibinfo {pages} {232--235} (\bibinfo {year}
  {2003})}\BibitemShut {NoStop}%
\bibitem [{\citenamefont {Roati}\ \emph {et~al.}(2007)\citenamefont {Roati},
  \citenamefont {Zaccanti}, \citenamefont {D'Errico}, \citenamefont {Catani},
  \citenamefont {Modugno}, \citenamefont {Simoni}, \citenamefont {Inguscio},\
  and\ \citenamefont {Modugno}}]{Roati_PhysRevLett.99.010403_2007}%
  \BibitemOpen
  \bibfield  {author} {\bibinfo {author} {\bibfnamefont {G.}~\bibnamefont
  {Roati}}, \bibinfo {author} {\bibfnamefont {M.}~\bibnamefont {Zaccanti}},
  \bibinfo {author} {\bibfnamefont {C.}~\bibnamefont {D'Errico}}, \bibinfo
  {author} {\bibfnamefont {J.}~\bibnamefont {Catani}}, \bibinfo {author}
  {\bibfnamefont {M.}~\bibnamefont {Modugno}}, \bibinfo {author} {\bibfnamefont
  {A.}~\bibnamefont {Simoni}}, \bibinfo {author} {\bibfnamefont
  {M.}~\bibnamefont {Inguscio}}, \ and\ \bibinfo {author} {\bibfnamefont
  {G.}~\bibnamefont {Modugno}},\ }\bibfield  {title} {\enquote {\bibinfo
  {title} {$^{39}\mathrm{K}$ bose-einstein condensate with tunable
  interactions},}\ }\href {\doibase 10.1103/PhysRevLett.99.010403} {\bibfield
  {journal} {\bibinfo  {journal} {Phys. Rev. Lett.}\ }\textbf {\bibinfo
  {volume} {99}},\ \bibinfo {pages} {010403} (\bibinfo {year}
  {2007})}\BibitemShut {NoStop}%
\bibitem [{\citenamefont {Dalfovo}\ \emph {et~al.}(1999)\citenamefont
  {Dalfovo}, \citenamefont {Giorgini}, \citenamefont {Pitaevskii},\ and\
  \citenamefont {Stringari}}]{Dalfovo_RevModPhys_1999}%
  \BibitemOpen
  \bibfield  {author} {\bibinfo {author} {\bibfnamefont {Franco}\ \bibnamefont
  {Dalfovo}}, \bibinfo {author} {\bibfnamefont {Stefano}\ \bibnamefont
  {Giorgini}}, \bibinfo {author} {\bibfnamefont {Lev~P.}\ \bibnamefont
  {Pitaevskii}}, \ and\ \bibinfo {author} {\bibfnamefont {Sandro}\ \bibnamefont
  {Stringari}},\ }\bibfield  {title} {\enquote {\bibinfo {title} {Theory of
  bose-einstein condensation in trapped gases},}\ }\href {\doibase
  10.1103/RevModPhys.71.463} {\bibfield  {journal} {\bibinfo  {journal} {Rev.
  Mod. Phys.}\ }\textbf {\bibinfo {volume} {71}},\ \bibinfo {pages} {463--512}
  (\bibinfo {year} {1999})}\BibitemShut {NoStop}%
\bibitem [{\citenamefont {Smerzi}\ and\ \citenamefont
  {Trombettoni}(2003)}]{Smerzi_PRA_2003}%
  \BibitemOpen
  \bibfield  {author} {\bibinfo {author} {\bibfnamefont {A.}~\bibnamefont
  {Smerzi}}\ and\ \bibinfo {author} {\bibfnamefont {A.}~\bibnamefont
  {Trombettoni}},\ }\bibfield  {title} {\enquote {\bibinfo {title} {Nonlinear
  tight-binding approximation for bose-einstein condensates in a lattice},}\
  }\href {\doibase 10.1103/PhysRevA.68.023613} {\bibfield  {journal} {\bibinfo
  {journal} {Phys. Rev. A}\ }\textbf {\bibinfo {volume} {68}},\ \bibinfo
  {pages} {023613} (\bibinfo {year} {2003})}\BibitemShut {NoStop}%
\end{thebibliography}%

\end{document}